# The roles of direct and environmental transmission in stochastic avian flu epidemic recurrence

May Anne E. Mata *[1,3], Priscilla E. Greenwood[2] and Rebecca C. Tyson[1]

[1] *University of British Columbia Okanagan, Kelowna, BC, Canada*
[2] *Department of Mathematics, University of British Columbia, Vancouver, BC Canada*
[3] *University of the Philippines Mindanao, Mintal, Davao City, Philippines*

**Abstract**

We present an analysis of an avian flu model that yields insight into the role of different transmission routes in the recurrence of avian influenza epidemics. Recent modelling work on avian influenza in wild bird populations takes into account demographic stochasticity and highlights the importance of environmental transmission in determining the outbreak periodicity, but only for a weak between-host transmission rate. We determine the relative contribution of environmental and direct transmission routes to the intensity of outbreaks. We use an approximation method to simulate noise sustained oscillations in a stochastic avian flu model with environmental and direct transmission routes. We see that the oscillations are governed by the product of a rotation and a slowly varying standard Ornstein-Uhlenbeck process (i.e., mean-reverting process). The intrinsic frequency of the damped deterministic version of the system predicts the dominant period of outbreaks. We show, using analytic computation of the intrinsic frequency and theoretical power spectral density, that the outbreak periodicity can be explained in terms of either or both types of transmission. The amplitude of outbreaks tends to be high when both types of transmission are strong.

*Corresponding author: mayang@alumni.ubc.ca

# 1 Introduction

Avian influenza is an infectious disease present in poultry and wild birds, and is known to pose threats to humans (World Health Organization, 2015). The disease pathogen is the avian influenza virus (AIV) whose natural hosts include aquatic birds (Krauss et al., 2004; Sharp et al., 1993) with wild ducks as its main reservoir (Kim and Negovetich, 2009). AIV strains can be either highly pathogenic (HP) or lowly pathogenic (LP) according to their ability to infect hosts. The HPAI viruses are the most virulent and are responsible for 'fowl plague' causing mortality as high as 100% in poultry (Alexander, 2000). The LPAI viruses are endemic in wild bird populations (Olsen et al., 2006; Webster and Bean, 1992) but can easily be transmitted to domestic stock and then mutate to HPAI type (Garamszegi and Møller, 2007). Regardless of the type of virus strain, the prevalence data for avian influenza displays recurrent epidemics over time. The underlying mechanism behind this outbreak pattern is the subject of active investigation (Breban and Drake, 2009; Clancy et al., 2006; Herrick et al., 2013; Roche and Lebarbenchon, 2009; Rohani et al., 2009), and is a key consideration in the development of effective control strategies for disease mitigation.

The virus spreads to healthy individuals either (i) by contact with an infected host i.e. through inter-host (direct) transmission, or (ii) by hosts acquiring the virus from the environment through drinking or filtering water while feeding, i.e., through environmental (indirect) transmission (Breban and Drake, 2009; Roche and Lebarbenchon, 2009). Recently, a few authors have highlighted the importance of environmental transmission as a driver of AIV epidemics. Rohani et al. (2009) demonstrated how neglecting environmental transmission could lead to underestimates for the explosiveness and duration of AIV epidemics. Breban et al. (2009) developed a new host-pathogen model combining within-season transmission dynamics, between-season migration and reproduction, and environmental variation, and showed that environmental transmission offers an explanation for the 2 to 4 year periodicity of AIV epidemics. Wang et al. (2012) formulated a simple stochastic model to show that increasing environmental transmission can make the outbreak period shorter. Their model predicted an outbreak period of 2 to 8 years. Wang et al. (2012) found this result consistent with the observed outbreak period obtained from wavelet analysis of empirical data (Krauss et al., 2004). Together, these papers and others point to environmental transmission as the key mechanism behind the approximate periodicity of AIV epidemics. They are based on the assumption that direct transmission is weak between wild birds. Recently, however, a simple SI model with only a direct transmission route, and without stochasticity, was found to provide the best fit to poultry outbreak data (Tuncer and Martcheva, 2013) suggesting that direct transmission may be stronger than originally thought. Motivated by these findings, we re-examine, mathematically, the contributions of the different transmission routes to the multi-year periodicity of avian flu epidemics.

The approximate multi-year periodicity of outbreaks is thought to be due to the random nature of contagion and recovery processes, together with demographic stochasticity (i.e., uncertainty in birth and death times), and the deterministic dynamics of the system (Nisbet and Gurney, 1982). It has been shown that demographic noise can sustain population oscillations that would otherwise damp to a stable equilibrium. These oscillations are commonly called noise sustained oscillations (Aparicio and Solari, 2001; Danøet al., 1999; Tomé and de Oliveira, 2009). Other treatments of this phenomenon in various contexts call it coherence resonance (Kuske et al., 2007), stochastic amplification (Alonso et al., 2007; McKane and Newman, 2005; McKane et al., 2007), or stochastic resonance (Dykman and McClintock, 1998; Gang; Mcnamara, 1989). Based on known AIV biology, a plausible model (Wang et al., 2012) suggests that AIV dynamics may arise from noise sustained oscillations.

A system exhibiting noise sustained oscillations can be analyzed using a recently developed approximation method. Baxendale and Greenwood (2011) showed that a stochastic process of two-dimensional noise sustained oscillations is, in distribution, approximately a rotation whose radius is modulated by a slowly varying bivariate standard Ornstein-Uhlenbeck (OU) process, a well-studied mean-reverting stochastic process (Uhlenbeck and Ornstein, 1930).

In this paper, we apply the approximation of Baxendale and Greenwood (2011) to a three-dimensional stochastic host-pathogen model, and assess the contributions of the direct and environmental transmission rates to the recurrent epidemics it produces. First, we show that the avian flu epidemic process can be approximated by the sum of a scaled univariate OU process and the product of a rotation and a bivariate slowly varying standard OU processes. Using the approximate stochastic process, we show that the outbreak period of the epidemics has a distribution centred at the intrinsic frequency of the associated deterministic part of the process, i.e., the deterministic analogue. We obtain the intrinsic frequency as a function of the two transmission rates and identify the relationship of each transmission rate with the dominant outbreak period. Furthermore, we determine how the outbreak intensity varies over a wide range of direct and environmental transmission rates.

We organize the paper as follows. In Section 2, we discuss the stochastic avian flu model formulated in terms of stochastic differential equations using the result of Kurtz (1978). In the same section, we also suggest an appropriate range of values to use for the transmission rate parameters. We review previous analytic work. Section 3 contains our own analysis, which includes determining the approximate process, the theoretical power spectral density (PSD), and the formula for the intrinsic frequency. We present in Section 4 the interpretation of our analysis, that the disease recurrence observed in stochastic simulations is approximately governed by a rotation matrix multiplied by a standard Ornstein-Uhlenbeck (OU) process. We then describe the influence of each transmission route on the dominant period of outbreaks (based on the intrinsic frequency) as well as the intensity of outbreaks (based on the stationary standard deviation of the approximate process). Finally, in Section 5, we discuss the implications of our results for characterizing avian flu epidemics and providing insights into the dependence of the frequency and variation of recurrent avian influenza epidemics on each transmission route.

## 2 The stochastic avian influenza model

### 2.1 Model description

Wang et al. (2012) formulated a stochastic model for avian influenza using the susceptible-infected-recovered (SIR) framework with an added environmental transmission rate. They found that their model was sufficient to explain the multi-year periodicity of flu outbreaks. In this host-pathogen model, there is a susceptible duck population of size $S$, an infected population of size $I$, a population of recovered individuals of size $R$, and an environmental virus concentration $V$. Note that $S$, $I$, and $R$ are integers, and all populations are functions of time. We also assume that a new susceptible duck is born once a host dies so that the total duck population is constant, i.e. $S+I+R = N$. Thus, we can solve for $R$ in terms of $S$ and $I$. By considering a short time interval $[t, t+\Delta t]$ and denoting by $T(\boldsymbol{\sigma}'|\boldsymbol{\sigma})$ the transition rate from state $\boldsymbol{\sigma} = (S, I, V)$ to $\boldsymbol{\sigma}' = \boldsymbol{\sigma} \pm \boldsymbol{\nu}$, where $\nu_i \in \{0, 1\}$, the different system events and their corresponding transition rates are:

1. Infection $$T(S-1, I+1, V|S, I, V) = \beta S \frac{I}{N} + \rho S \frac{V}{N_V} \tag{1a}$$

2. Birth and Death $$T(S+1, I, V|S, I, V) = \mu(N - S - I), \tag{1b}$$
$$T(S, I+1, V|S, I, V) = \mu I, \tag{1c}$$
$$T(S, I, V+1|S, I, V) = \tau I + \delta V, \tag{1d}$$
$$T(S, I, V-1|S, I, V) = \eta V. \tag{1e}$$

3. Recovery $$T(S, I-1, V|S, I, V) = \gamma I. \tag{1f}$$

We schematically present these processes in Figure 1.

The first event (1a) describes the infection of a susceptible individual. Infection happens when a susceptible host is in close contact with an infected host or when it acquires the virus directly from the environment. The rate of transmission of the disease is given by the likelihood of contact between a susceptible individual and either infected individuals or virions in the environment, multiplied by the rate at which virions are acquired by susceptible individuals in each case. Note that the likelihood of contact depends on the fraction of infected individuals ($I/N$) and the concentration of virus in the environment normalised by a reference concentration ($V/N_V$), i.e., the likelihood of contact in a frequency-dependent process.

The second category of events (1b)-(1e) encompasses the birth and death processes of the host and the virus. For simplicity, it is assumed here that the per capita host birth and death rates have the same value $\mu$. We assume that virus is introduced into the environment at a constant rate $\delta$ from alternative hosts. The virus concentration in the environment also grows when infected ducks shed virions; this shedding occurs at rate $\tau$. The clearance rate of the virus in the environment is $\eta$.

Finally, the third category of events (1f) is the recovery of infected ducks at per capita rate $\gamma$.

The parameters $\beta$, $\mu$, $\tau$, $\delta$, $\eta$ and $\gamma$ are stochastic rates.

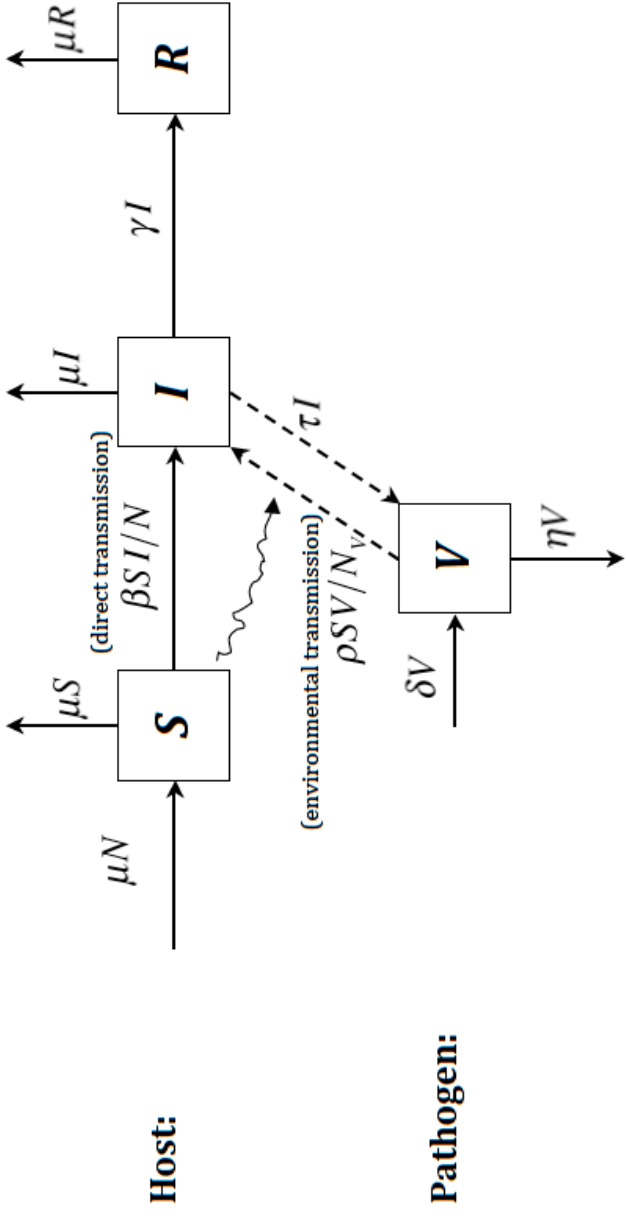

**Figure 1:** A schematic diagram of the host-pathogen model for avian influenza adapted from Wang et al. (2012). The host population has three compartments corresponding to susceptible, infected, and removed individuals. The solid arrows represent movement of individuals from one compartment to another as a result of birth, death, infection, and recovery processes. The dashed arrows represent the increase or decrease in the number of infected individuals due to the interaction (squiggly arrow) of the susceptible host with the avian influenza virus present in the environment.

The resulting stochastic avian flu host-pathogen model (see Appendix A) is approximated for large $N$ by the following system of stochastic differential equations (SDEs):

$$ds = (-\beta si - \rho sv + \mu(1-s))\,dt + \frac{1}{\sqrt{N}}\left(-G_1 dW_1 + G_2 dW_2 + G_3 dW_3\right),$$
$$di = (\beta si + \rho sv - (\mu+\gamma)i)\,dt + \frac{1}{\sqrt{N}}\left(G_1 dW_1 - G_3 dW_3 - G_4 dW_4\right), \quad (2)$$
$$dv = (k\tau i + \delta v - \eta v)\,dt + \frac{1}{\sqrt{N_V}}\left(G_5 dW_5 - G_6 dW_6\right),$$

where,

$$G_1 = \sqrt{\beta si + \rho sv}, \quad G_2 = \sqrt{\mu(1-s-i)}, \quad G_3 = \sqrt{\mu i},$$
$$G_4 = \sqrt{\gamma i}, \quad G_5 = \sqrt{k\tau i + \delta v}, \quad \text{and} \quad G_6 = \sqrt{\eta v}. \quad (3)$$

Here, $s = S/N, i = I/N, v = V/N_V$, and $k = N/N_V$. The SDE for the proportion of recovered ducks $r(t)$ is not necessary because we eliminate $r(t)$ using $r(t) = 1 - s(t) - i(t)$, which follows from $S + I + R = N$. In (2), the second term vanishes as $N, N_V \to \infty$ which leads us to the deterministic, or so-called mean-field, dynamics as found in Wang et al. (2012):

$$\dot{\phi}_1 = -\beta\phi_1\phi_2 - \rho\phi_1\psi + \mu(1-\phi_1),$$
$$\dot{\phi}_2 = \beta\phi_1\phi_2 + \rho\phi_1\psi - (\mu+\gamma)\phi_2, \quad (4)$$
$$\dot{\psi} = \kappa\tau\phi_2 + \delta\psi - \eta\psi.$$

The deterministic variables $\phi_1, \phi_2, \psi$ represent fractions of the susceptible hosts, infected hosts, and virus in the environment, respectively, while $\kappa = \lim_{N,N_V \to \infty} \frac{N}{N_V}$.

The basic reproduction number $\mathcal{R}_0$, i.e., the average number of secondary infections resulting from one infected individual in a susceptible population, is given by (Wang et al., 2012),

$$\mathcal{R}_0 = \frac{\beta}{\mu+\gamma} + \frac{\kappa\rho\tau}{(\eta-\delta)(\mu+\gamma)}. \quad (5)$$

The deterministic system has a stable endemic equilibrium, which can be written in terms of $\mathcal{R}_0$,

$$(\phi_1^*, \phi_2^*, \psi^*) = \left(1/\mathcal{R}_0, \frac{\mu}{\mu+\gamma}(1-1/\mathcal{R}_0), \frac{\kappa\mu\tau}{(\eta-\delta)(\mu+\gamma)}(1-1/\mathcal{R}_0)\right). \quad (6)$$

When the basic reproductive number $\mathcal{R}_0 > 1$, the disease is epidemic.

We are interested in characterizing the fluctuations of the stochastic model around the steady-state solution. Hence, we linearize (2) around $(\phi_1^*, \phi_2^*, \psi^*)$ and obtain the linear diffusion equation (see Appendix B for the detailed derivation),

$$d\boldsymbol{\xi} = \mathbf{A_0}\boldsymbol{\xi}\,dt + \mathbf{C_0}\,d\mathbf{W}, \quad \boldsymbol{\xi}(t), \mathbf{W}(t) \in \mathbb{R}^3, \mathbf{A_0}, \mathbf{C_0} \in \mathbb{R}^{3\times 3}. \quad (7)$$

where

$$\mathbf{A_0} = \begin{bmatrix} -\beta\phi_2^* - \rho\psi^* - \mu & -\beta\phi_1^* & -\rho\phi_1^* \\ -\beta\phi_2^* & \beta\phi_1^* - \mu - \gamma & \rho\phi_1^* \\ 0 & \kappa\tau & \delta - \eta \end{bmatrix}, \quad (8)$$

and

$$\begin{aligned}
\mathbf{C_0} &= \begin{bmatrix} C_{11} & C_{12} & 0 \\ C_{21} & C_{22} & 0 \\ 0 & 0 & C_{33} \end{bmatrix}^{1/2}, \\
C_{11} &= \beta \phi_1^* \phi_2^* + \rho \phi_1^* \psi^* + \mu(1 - \phi_1^*), \\
C_{12} &= C_{21} = -\beta \phi_1^* \phi_2^* - \rho \phi_1^* \psi^* - \mu \phi_2^*, \\
C_{22} &= \beta \phi_1^* \phi_2^* + \rho \phi_1^* \psi^* + (\mu + \gamma) \phi_2^*, \\
C_{33} &= \kappa \tau \phi_2^* + \delta \psi^* + \eta \psi^*.
\end{aligned} \quad (9)$$

Equations (7)-(9) are the Langevin equations derived by Wang et al. (2012). The drift coefficient matrix $\mathbf{A_0}$ is the Jacobian of the deterministic model (4) evaluated at the endemic equilibrium $(\phi_1^*, \phi_2^*, \psi^*)$. The diffusion matrix $\mathbf{C_0}$ is formed using the coefficients of the independent Wiener processes in (2). This matrix is the square-root of the covariance matrix $\mathcal{B}$ found by Wang et al. (2012). Our paper focusses on the study of the linear system (7).

## 2.2 Parameter values

All simulations produced in this work use as default parameters the values in Table 1 largely taken from Wang et al. (2012). The values of parameters $\mu$, $\eta$, $\beta$, and $\gamma$ are based on empirical studies in the literature (see caption of Table 1). No data are available for $\rho$ and $\delta$ and so they are varied within the range used by Wang et al. (2012). Note that the unit of the shedding rate $\tau$ is virion/mL/duck/year rather than virion/duck/day as erroneously reported in Wang et al. (2012).

The values for the transmission parameters $\beta$ and $\rho$ can vary widely with seasonal climate changes and from one geographic area to another. We thus give a range of values for $\beta$ and $\rho$ and study the system's response to different levels and modes of transmission. As displayed in Table 1, we use a wider range of $\beta$ values than was used by Wang et al. (2012). Observe that the infection term in (4) is given by $\beta \phi_2 = \beta I/N$ (rather than simply $\beta I$). In ecological terms, this means that $\beta$ is the transmission rate for a frequency-dependent process rather than a density-dependent process. Roche and Lebarbenchon (2009) find that $\beta$ ranges from 0.00005 to 1500. In this paper, we plot our results for $\beta$ values ranging from 0 to 300.

Avian influenza epidemic models exhibit noise sustained oscillations with a nonzero dominant frequency for a small range of $\beta$ values, with some fixed value of $\rho$. Knowing only that $\beta$ falls within a wide range of values, re-investigation of the relative contribution of each of the transmission modes is necessary. In particular, a larger $\beta$ poses the possibility that direct transmission ($\beta$) alone, along with stochasticity, can drive avian flu epidemics with multi-annual periodicity.

**Table 1:** Descriptions of the parameters in the avian flu model and the values used in stochastic simulations. The range of values for $\beta$ in bold-faced is the set of values that were used in (Wang et al., 2012).

| Parameter | Value/Range | Description | Unit |
|---|---|---|---|
| $N$ | $10^3$ | Duck population size | duck |
| $N_V$ | $10^5$ | Virus concentration in the environment | virion/mL |
| $\mu$ | 0.3 | Duck birth and death rate | year$^{-1}$ |
| $\delta$ | 0.1 | Virus replication rate | year$^{-1}$ |
| $\tau$ | $10^4$ | Virus shedding rate | virion mL$^{-1}$/duck/year |
| $\eta$ | 3 | Virus clearance rate | year$^{-1}$ |
| $\beta$ | 0.00005 to 1500 **(0-0.05)** | Direct transmission rate | duck$^{-1}$year$^{-1}$ |
| $\rho$ | $0-3$ | Environmental transmission rate | year$^{-1}$ |
| $\gamma$ | 5.5 | Duck recovery rate | year$^{-1}$ |

Assessing the relative contribution of each of the transmission rates to disease recurrence requires a quantitative comparison of results. From (5), we see that if both

$$\beta < \mu + \gamma \quad \text{and} \quad \rho < \frac{(\eta - \delta)(\mu + \gamma)}{\kappa \tau}, \tag{10}$$

then $\mathcal{R}_0 < 1$. Using the values in Table 1, we find that when both $\beta < 5.8$ and $\rho < 0.16$ we have $\mathcal{R}_0 < 1$.

## 2.3 Preliminary analysis of the model

In this section we highlight key results from the analysis of the model of Wang et al. (2012), and extend these results to include the role of $\mathcal{R}_0$ in the dynamics of avian flu. It is an established fact that the oscillations of a damped system can be sustained by stochasticity. Wang et al. (2012) showed that the deterministic version of model (4) has a stable endemic steady-state (6) when $\mathcal{R}_0 > 1$. Here, we take the calculation one step further by determining the parameter ranges where (6) is a stable sink or a stable focus. We do this by plotting the eigenvalues of the Jacobian in (8) against $\mathcal{R}_0$ from (5). Figure 2 displays this plot.

In Figure 2, we see the eigenvalues of $\mathbf{A}_0$ plotted as functions of $\mathcal{R}_0$. We use the parameter values in Table 1 with $\beta = 0.05$. As $\rho$ increases, so does $\mathcal{R}_0$. We observe that for $\mathcal{R}_0 < 1$, all of the eigenvalues are real and one has positive sign. As expected from Wang et al. (2012), the endemic steady-state is unstable for this case and the system evolves to the disease-free equilibrium. A transition in the signs of the eigenvalues happens at $\mathcal{R}_0 = 1$. When $\mathcal{R}_0 > 1$, the system gives rise to a complex-conjugate pair of eigenvalues with negative real parts, and a negative real eigenvalue. Thus, the steady-state $(\phi_1^*, \phi_2^*, \psi^*)$ is a stable focus, i.e., the deterministic system exhibits damped population cycles. Note that a similar result is obtained if we fix $\rho$ and let $\mathcal{R}_0$ vary with $\beta$.

Suppose that we represent the complex eigenvalues as $-\lambda \pm i\omega$ where $\lambda$ and $\omega$ are magnitudes of the real and imaginary parts, respectively. For $\mathcal{R}_0 > 1$ in Figure 2, $\lambda$ is clearly smaller than $\omega$, because the upper solid curve is above the lower dashed curve with both curves drawn below the $x$-axis. Moreover, observe that $\omega$ increases faster than $\lambda$. Hence we deduce that the ratio $\lambda/\omega$ decreases as $\mathcal{R}_0$ increases.

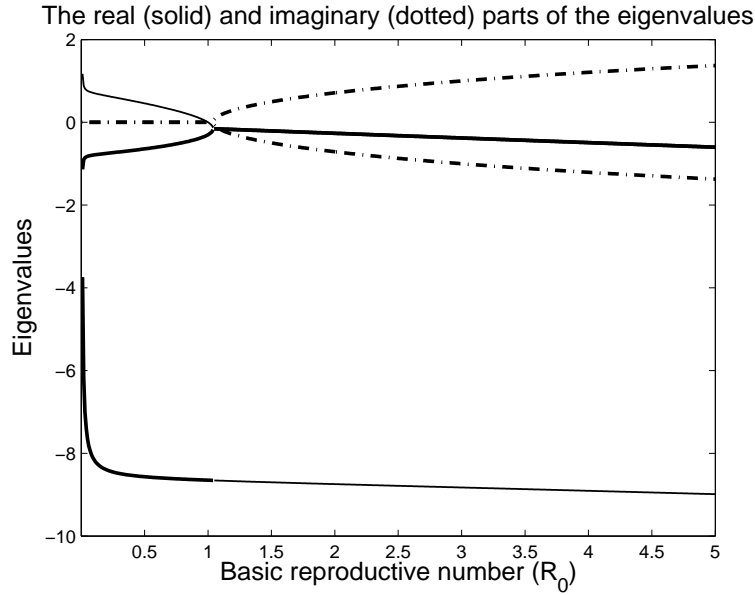

**Figure 2:** The real (solid curves) and imaginary (dash-dot curves) parts of the eigenvalues of $\mathbf{A}_0$ in (8) associated with the stability of the endemic steady-state (6) plotted against $\mathcal{R}_0$ in the case where $\beta = 0.05$. There are three real eigenvalues when $\mathcal{R}_0 < 1$, of which two are negative (thick solid lines). For $\mathcal{R}_0 > 1$, we have a complex-conjugate pair of eigenvalues with negative real parts (thick solid lines) and another negative eigenvalue (thin solid line). The imaginary parts of the complex eigenvalues are shown by the dotted lines. Other parameter values are shown in Table 1.

In Figure 3, we plot typical stochastic realizations of the avian flu model (2) for $\mathcal{R}_0 > 1$ using various combinations of parameters. These stochastic paths display oscillations sustained by noise.

We display the plots of stochastic realizations for the case when $\beta = 0.05$ but with different $\rho$ values in Figures 3(a) and 3(b). Here we notice that higher amplitude and higher frequency of epidemics are observed in Figure 3(b) as compared to the fluctuation plotted in Figure 3(a) where $\rho$ is twice as much. This observation is consistent with Figure 2.

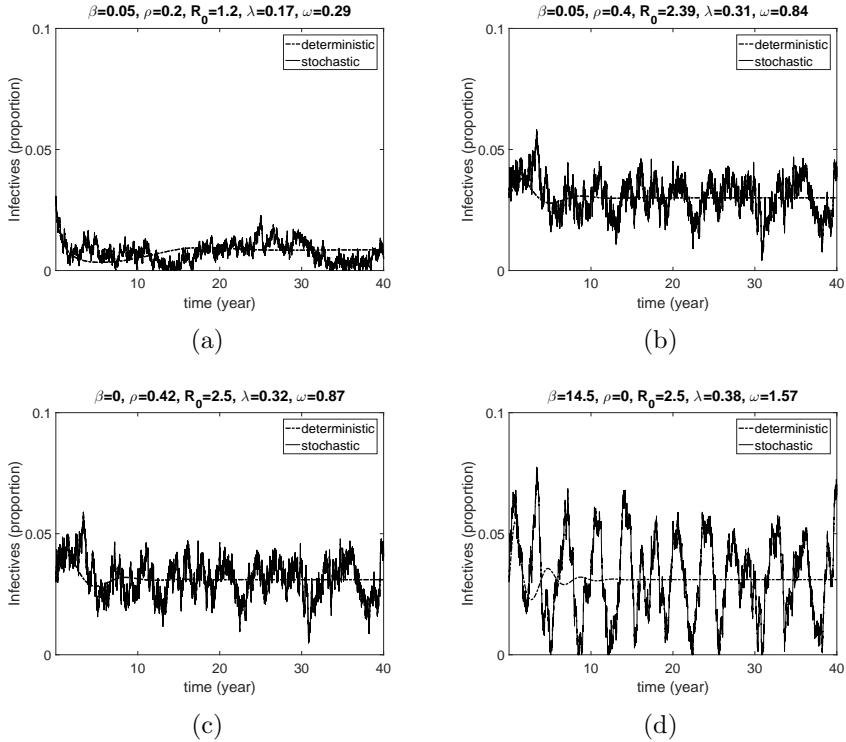

**Figure 3:** Simulation of the stochastic model (2) and its corresponding deterministic solution for $\beta = 0.05$ with (a) $\mathcal{R}_0 = 1.2$ and (b) $\mathcal{R}_0 = 2.5$, and for $\mathcal{R}_0 = 2.5$ with (c) $\beta = 0$ ($\rho = 0.4205$) and (d) $\rho = 0$ ($\beta = 14.5$).

In Figures 3(c) and 3(d), we plot stochastic paths for the case when either $\beta = 0$ or $\rho = 0$ but with roughly the same $\mathcal{R}_0$. In this comparison, $\mathcal{R}_0 \approx 2.5$, the upper bound of the confidence interval for the estimate of $\mathcal{R}_0$ for avian influenza in wild birds (van der Goot et al., 2003). Comparing Figures 3(c) and 3(d), we observe that the periodicity and intensity of outbreaks in the two cases is different, even though the basic reproduction number is the same in the two cases. Current theory points toward the transmission mode as a determining factor for understanding the periodicity and intensity of avian flu outbreaks. Here, we develop a method that allows us to determine mathematically the effect of each transmission mode on the recurrence (periodicity and intensity) of avian influenza.

## 3 Analytic methods

In this section we develop the analytic tools we need to understand the contribution of each transmission route to the dynamic behaviour of the model. In particular, our goal is to develop a mathematical description of the noise sustained oscillations that are observed in stochastic simulations, such as those shown in Figure 3.

## 3.1 Approximate solutions

We begin by defining an approximation for the process of oscillations produced by the stochastic avian flu model. Our starting point is the linear diffusion equation given by (7), which describes the process $\boldsymbol{\xi} = (\xi_S, \xi_I, \xi_V)$ near the endemic steady-state for large time. Under certain conditions, an approximate solution of (7) can be obtained using an extension of a result from Baxendale and Greenwood (2011). Specifically, if the eigenvalues of the drift coefficient matrix $\mathbf{A}_0$ (see (68) below) are of the form $-\zeta$, and $-\lambda \pm i\omega$ with $\lambda/\omega$ small, for $\zeta, \lambda, \omega > 0$, an approximate solution for (7) (see Appendix C for details) is given by

$$\boldsymbol{\xi}^{app}(t) = y_1(t)\mathbf{Q}_{\bullet 1} + \frac{\tilde{\sigma}}{\sqrt{\lambda}}[\mathbf{Q}_{\bullet 2}, \mathbf{Q}_{\bullet 3}]\mathbf{R}_{-\omega t}\mathbf{S}_{\lambda t}, \qquad (11)$$

where $\mathbf{Q}$ is the canonical form of the eigenvector matrix associated with the three eigenvalues. The vector $\mathbf{Q}_{\bullet \mathbf{j}}$ denotes the $j$th column of $\mathbf{Q}$. The stochastic process $y_1(t)$ is an Ornstein-Uhlenbeck (OU) process (Uhlenbeck and Ornstein, 1930) with mean zero, decay rate $\zeta$, and diffusion coefficient $\sigma_1$ obtained using the first row of the matrix $\mathbf{Q}^{-1}\mathbf{C_0}$. The matrix $\mathbf{R}_{-\omega t}$ is a rotation matrix. It describes the circular motion of the process with frequency $\omega$. The vector process $\mathbf{S}_{\lambda t}$ is a bi-variate OU process with independent components. The scalar $\tilde{\sigma}$ is determined by the last two rows of $\mathbf{Q}^{-1}\mathbf{C_0}$ (the full expression is given by (49) in Appendix C). We express the large-time stationary solution of (2) as:

$$\begin{aligned} s(t) &= \phi_1^* + \frac{1}{\sqrt{N}}\xi_S(t) \approx \phi_1^* + \frac{1}{\sqrt{N}}\xi_S^{app}(t), \\ i(t) &= \phi_2^* + \frac{1}{\sqrt{N}}\xi_I(t) \approx \phi_2^* + \frac{1}{\sqrt{N}}\xi_I^{app}(t), \\ v(t) &= \psi^* + \frac{1}{\sqrt{N_V}}\xi_V(t) \approx \psi^* + \frac{1}{\sqrt{N_V}}\xi_V^{app}(t). \end{aligned} \qquad (12)$$

The formulation (12) shows that the solution of (2) near the endemic steady-state behaves approximately like the product of a rotation and an OU process (11).

## 3.2 Power spectral density

We use the theoretical power spectral density (PSD) to determine the distribution of frequency components within the stochastic process produced by (7) with (8) and (9), and by their approximate form (11). A linear diffusion process is described as a general multivariate OU process, i.e.,

$$d\mathbf{x} = -\mathbf{A}\mathbf{x}(t)\, dt + \mathbf{B}\, d\mathbf{W}(t). \qquad (13)$$

According to Gardiner (1986), the PSD of an $n$-dimensional process is obtained from the main diagonal of the matrix given by:

$$\mathbf{S}(f) = \frac{1}{2\pi}(\mathbf{A} + if)^{-1}\mathbf{B}\mathbf{B}^T(\mathbf{A}^T - if)^{-1}. \qquad (14)$$

From (14), it follows that the PSD of $\boldsymbol{\xi}(t)$ satisfying (7) is obtained from the main diagonal of the matrix

$$\mathbf{S}(f) = \frac{1}{2\pi}(-\mathbf{A}_0 + if)^{-1}\mathbf{C}_0\mathbf{C}_0^T(-\mathbf{A}_0^T - if)^{-1}. \qquad (15)$$

On the other hand, the theoretical PSD of the approximate solution $\boldsymbol{\xi}^{app}(t)$ in (11) is determined using the coefficients of (see Appendix C for derivation)

$$d\boldsymbol{\xi}^{app}(t) = \mathbf{A_0}\boldsymbol{\xi}^{app}\ dt + \mathbf{Q}\cdot\text{diag}(\sigma_1, \tilde{\sigma}, \tilde{\sigma})\ d\mathbf{W}(t).$$

Thus, the PSD of $\boldsymbol{\xi}^{app}$ is obtained from

$$\mathbf{S}(f) = \frac{1}{2\pi}\cdot\text{diag}(\sigma_1^2, \tilde{\sigma}^2, \tilde{\sigma}^2)\cdot(-\mathbf{A}_0 + if)^{-1}\mathbf{Q}\mathbf{Q}^T(-\mathbf{A}_0^T - if)^{-1}. \tag{16}$$

## 3.3 The intrinsic frequency and the decay rate of the deterministic dynamics

The approximation (11) depends explicitly on the intrinsic frequency $\omega$ and the decay rate $\lambda$ of the damped oscillations predicted by the deterministic system (4). We can derive these quantities by obtaining the eigenvalues of the Jacobian matrix $\mathbf{A}_0$, which are the solutions of the characteristic polynomial given by

$$\nu^3 - a\nu^2 - b\nu - c = 0, \tag{17}$$

where:
$$\begin{aligned}
a &= (\delta - \eta) - \mu\mathcal{R}_0 - \gamma - \mu + \beta/\mathcal{R}_0, \\
b &= -\mu(\eta - \delta + \gamma + \mu)\mathcal{R}_0 + \mu\beta/\mathcal{R}_0, \\
c &= -\mu(\eta - \delta)(\gamma + \mu)(\mathcal{R}_0 - 1).
\end{aligned} \tag{18}$$

In Appendix F, we derive $\lambda$ and $\omega$ as functions of $\beta$ and $\rho$, using the roots of (17) and the avian flu parameters in Table 1. The results are shown in Figures 6 and 7 where we see $\lambda/\omega$ and $\omega$ are functions of $\beta$ and $\rho$.

## 3.4 Numerical tools and functions

All numerical computations were done using MATLAB (2010). We computed the solution of the deterministic avian flu model (4) with the built-in function ode45(), an ordinary differential equation solver that is based on the Runge-Kutta (4,5) formula. Stochastic simulations of (2) and (7) were done using the Euler scheme (or the Euler-Maruyama scheme) (Øksendal, 2003), a first-order discretization method for stochastic differential equations. A time step of 0.01 was used for simulations of (2), (7), and (4).

# 4 Results

We present our results in two parts. First, we substitute the parameter values from Table 1 into the approximate solution (11) of (7) with (8) and (9). Second, we use the approximation to understand how the different transmission routes affect the dominant periodicity and the typical intensity of epidemics (see Section 4.2).

## 4.1 An approximate avian flu epidemic process

In this section, we use the approximation (11) to describe explicitly the noise sustained oscillations we saw in Figure 3. For consistency with previous work (Wang et al., 2012), we have chosen $\beta = 0.05$ and $\rho = 0.4$. This choice of parameters results in the endemic steady-state $\phi_1^* = 0.419$, $\phi_2^* = 0.03$, $\psi^* = 1.04$ with $\mathcal{R}_0 \approx 2.39$. The deterministic process defined by (4) persists. The diffusion equation (7) with (8) and (9) becomes

$$d\boldsymbol{\xi} = \mathbf{A}_0 \boldsymbol{\xi}\, dt + \mathbf{C}_0\, d\mathbf{W}, \qquad \text{where}$$
$$\mathbf{A}_0 = \begin{bmatrix} -0.716 & -0.021 & -0.168 \\ 0.410 & -5.78 & 0.168 \\ 0 & 100 & -2.9 \end{bmatrix}, \qquad \text{and}$$
$$\mathbf{C}_0 = \begin{bmatrix} 0.568 & -0.161 & 0 \\ -0.161 & 0.568 & 0 \\ 0 & 0 & 2.494 \end{bmatrix} \tag{19}$$

The eigenvalues of the Jacobian $\mathbf{A}_0$ in (19) are $-\zeta = -8.78$ and $-\lambda \pm i\omega = -0.309 \pm 0.838i$. The ratio $\lambda/\omega = \frac{0.309}{0.838} \approx 0.369$ is sufficiently small that we can approximate the process $\boldsymbol{\xi}(t)$ by (11), as evidenced by the comparison of PSDs in Figure 4.

Using the eigenvalues above, we arrive at the canonical matrix of eigenvectors, i.e., the columns are eigenvectors of $\mathbf{A}_0$, given by

$$\mathbf{Q} = \begin{bmatrix} 0.021 & -0.078 & 0.16 \\ -0.059 & 0.026 & 0.008 \\ 0.998 & 0.984 & 0 \end{bmatrix}. \tag{20}$$

The matrix $\mathbf{C}_0$ in the diffusion term of (19) gives us

$$\boldsymbol{\Sigma} = \mathbf{Q}^{-1}\mathbf{C}_0 = \begin{bmatrix} 2.13 & -6.43 & 0.834 \\ -2.16 & 6.52 & 1.69 \\ 2.23 & 3 & 0.715 \end{bmatrix}. \tag{21}$$

We compute $\sigma_1$ (see Appendix C) by taking the norm of the first row of the matrix $\boldsymbol{\Sigma}$ in (21),

$$\sigma_1 = ||(2.13, -6.43, 0.834)|| \approx 6.82.$$

We form the matrix $\tilde{\mathbf{C}} = (\tilde{\boldsymbol{\Sigma}}\tilde{\boldsymbol{\Sigma}}^\intercal)^{1/2}$ where,

$$\tilde{\boldsymbol{\Sigma}} = \begin{bmatrix} -2.16 & 6.52 & 1.69 \\ 2.23 & 3 & 0.715 \end{bmatrix}, \tag{22}$$

to obtain $\tilde{\sigma}^2 = Tr(\tilde{\mathbf{C}}\tilde{\mathbf{C}}^\intercal)/2$ and so $\tilde{\sigma} \approx 5.68$.

Using (11), the solution to (19) near the endemic steady-state is approximately

$$\begin{bmatrix} \xi_S^{app}(t) \\ \xi_I^{app}(t) \\ \xi_V^{app}(t) \end{bmatrix} = y_1(t) \begin{bmatrix} 0.021 \\ -0.059 \\ 0.998 \end{bmatrix}$$
$$+ 10.2 \begin{bmatrix} -0.078 & 0.16 \\ 0.026 & 0.008 \\ 0.983 & 0 \end{bmatrix} \mathbf{R}_{-0.838t} \mathbf{S}_{0.309t}, \tag{23}$$

where
$$dy_1 = -8.78 y_1 \, dt + 6.82 \, dW_1. \tag{24}$$

One way to see if the process (23) is a reasonable approximate solution for (19) is to compute the theoretical PSDs of the exact solution $\boldsymbol{\xi}(t)$ of (19) and the approximation $\boldsymbol{\xi}^{app}(t)$ in (23) using formulae (15) and (16), respectively. In Figure 4, the PSDs of fluctuations $\boldsymbol{\xi}(t)$ and $\boldsymbol{\xi}^{app}(t)$ agree fairly well. The dominant frequency is close to the intrinsic frequency $\omega = 0.8377$.

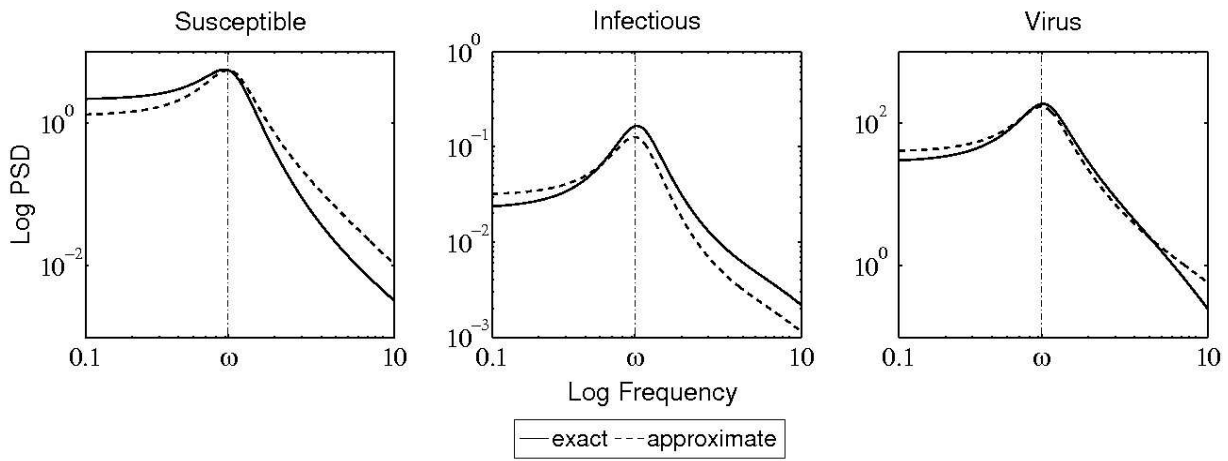

**Figure 4:** Comparisons between the theoretical PSD of the exact process $\boldsymbol{\xi}(t)$ (solid line) satisfying (19) and the approximate process $\boldsymbol{\xi}^{app}(t)$ (dashed line) given by (23), for the fluctuations of the susceptible, infectious, and the virus populations. Default parameter values are in Table 1 with $\beta = 0.05$ and $\rho = 0.4$.

The process $\boldsymbol{\xi}^{app}(t)$ depends on the OU process $y_1(t)$ whose dynamics are described by (24), an OU process with asymptotic mean zero, decay rate 8.78, and diffusion coefficient 6.82. From the first term of (23), we find that the contribution of the process $y_1(t)$ to host fluctuation processes, $\xi_S(t)$ and $\xi_I(t)$, is very small as compared to that of the virus fluctuation process $\xi_V(t)$. We use the stationary standard deviation to measure the typical amplitude of the population fluctuations. Note that the stationary variance of the process $y_1(t)$ is $46.51/(2 \times 8.78) \approx 2.65$, i.e. stationary standard deviation $\sqrt{2.65} \approx 1.63$. The stationary standard deviation, i.e. typical amplitude, of the stochastic path for the process $\boldsymbol{\xi}^{app}(t)$ (see (27) or (57) in Appendix C), is given by

$$SSD_i \approx \sqrt{2.65 q_{i1}^2 + 10.2^2 r_i^2}. \tag{25}$$

The computed typical amplitudes of the susceptible, infectious, and virus population fluctuations are $1.81, 0.29$, and $10.18$, respectively. In Figure 5, the stochastic realizations for the different population fluctuations given by (19) and (11) are plotted along with their stationary standard deviations indicated by horizontal lines to indicate that the stochastic oscillations generally lie within one stationary standard deviation.

Writing (23) in polar form (see Appendix C), we obtain

$$\begin{aligned}
\xi_S(t) &\approx 0.021 y_1(t) + 1.82|\mathbf{S}_{0.309t}|\cos(\varphi_{0.309t} - 0.838t - 2.02), \\
\xi_I(t) &\approx -0.059 y_1(t) + 0.275|\mathbf{S}_{0.309t}|\cos(\varphi_{0.309t} - 0.838t - 0.313), \\
\xi_V(t) &\approx 0.998 y_1(t) + 10|\mathbf{S}_{0.309t}|\cos(\varphi_{0.309t} - 0.838t).
\end{aligned} \qquad (26)$$

We know from the approximate form (26) that the phase differences between the population fluctuation processes $\xi_i(t)$ are constants. For instance, the susceptible duck population fluctuation is out of phase with other populations in the system. With respect to the virus population, the susceptible duck population exhibits noise-sustained oscillations with a phase advance, with respect to the virus population, of approximately $2.02/\omega = 2.02/0.838 \approx 2.4$ years. The infectious duck population, on the other hand, oscillates with a phase advance of $0.313/0.838 \approx 0.37$ years from the virus population.

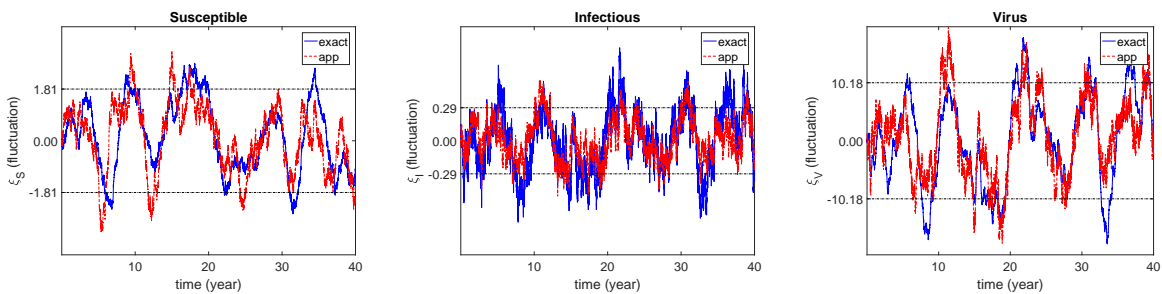

**Figure 5:** A stochastic realization of the population fluctuations by simulating (19) (solid line) and (23) (dashed line) and their corresponding stationary standard deviations, i.e. typical amplitudes (in horizontal lines) computed using (26).

## 4.2 The relative contribution of the different transmission routes

One of the key factors leading to outbreaks is the rate of transmission. As outlined previously, there are two transmission routes for avian flu: direct transmission of virus from one duck to another, and indirect transmission of viruses via the environment. We use the approximation $\boldsymbol{\xi}^{app}(t)$ to shed light on the relative importance of direct and indirect transmission to outbreak occurrence.

As a first step, we identify the constraints for $\beta$ and $\rho$ under which one expects to observe recurrent epidemics and where the approximation is valid. We use the parameter ranges from Table 1 and display in the top panels of Figure 6 a plot of $\lambda/\omega$ as a function of $\beta$ and $\rho$. In Figure 6(a), we see a triangular region surrounding $\beta = \rho = 0$ where the approximation is no longer relevant because no recurrent epidemics can be observed there. In the triangular parameter region, $\mathcal{R}_0 \leq 1$ as seen in Figure 6(c) and there are no recurrent epidemics. Within a large portion of the remaining parameter region defined by $0 \leq \beta \leq 15$ and $0 \leq \rho \leq 1$ (see Figure 6(a)), $\lambda/\omega$ is sufficiently small that the approximation is valid.

We further extend the range of transmission parameters to $0 \leq \beta \leq 300$ and $0 \leq \rho \leq 3$ as displayed in Figure 6(b) to show the behaviour of $\lambda/\omega$ for larger values of $\beta$ and $\rho$. We observe that when $\beta$ and $\rho$ are both very large, as depicted by the white region in the upper portion of Figure 6(b), $\lambda/\omega > 1$ and so the approximation is not necessarily valid. In Figure 6(b), a black curve is drawn, which corresponds to $\lambda/\omega = 1$. The approximation is valid for $\beta$ and $\rho$ values in the region well below this curve. We will focus on the range $0 \leq \rho \leq 1$ and $0 \leq \beta \leq 100$. The following analysis focusses on this region to study how the different routes of transmission influence the periodicity and intensity of recurrent epidemics reflected by the model.

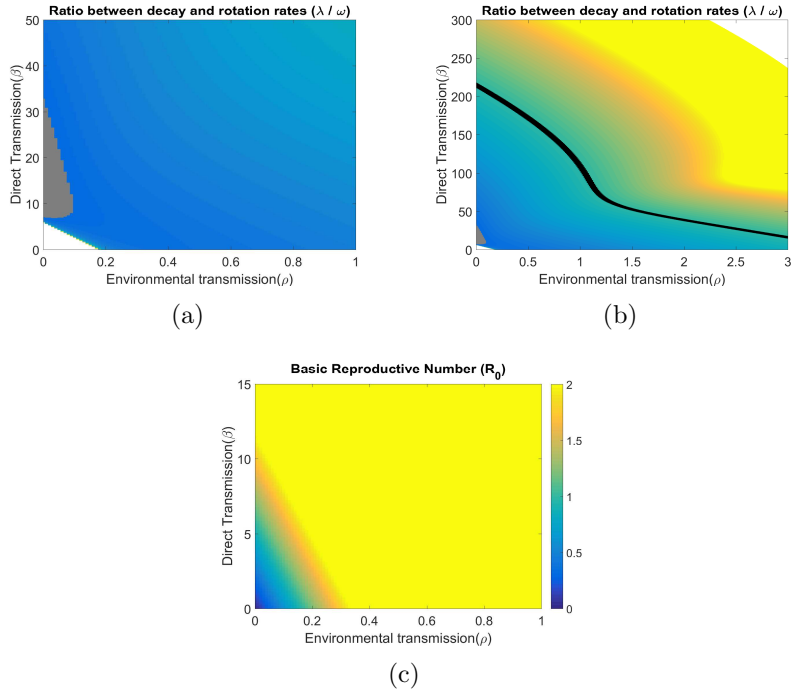

**Figure 6:** (a,b) Plot of the ratio $\lambda/\omega$ as a function of $\beta$ and $\rho$. The triangular white region in the lower left is where no recurrence is observed. Panel (a) considers $0 \leq \beta \leq 50$ and $0 \leq \rho \leq 1$ while Panel (b) considers $0 \leq \beta \leq 300$ and $0 \leq \rho \leq 3$. The grey region in both panels is where $\lambda/\omega \leq 0.35$. In Panel (b), the black curve corresponds to $\lambda/\omega = 1$. (c) Plot of the basic reproduction number $\mathcal{R}_0$ as a function of $\beta$ and $\rho$. The darker (blue online) shade indicates that no epidemic can be observed in the model for this parameter region. Parameter values and ranges from the literature are in Table 1.

### 4.2.1 Dominant outbreak period

The period of any oscillating function is the inverse of the frequency. We analyze the dominant outbreak period of the simulated epidemic from (7) by considering the intrinsic frequency of the deterministic system. For a parameter range where the approximation (11) is close to the exact process satisfying (7) (see Figure 4), the dominant frequencies predicted by the two are very close. In addition, this frequency is close to that from the deterministic system when $\lambda$ is small (Greenwood et al., 2014). Hence we use $\omega$ given by formula (71) (see Appendix F).

Using formula (71) in Appendix F, we compute the intrinsic frequency $\omega = \omega(\rho, \beta)$ over $0 \leq \rho \leq 1$ and $0 \leq \beta \leq 100$ for the parameters in Table 1 to obtain Figure 7.

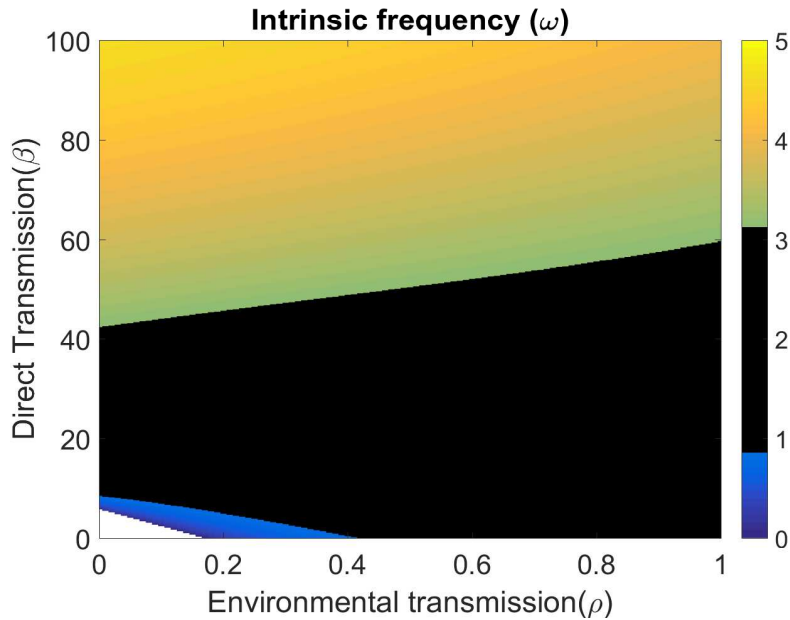

**Figure 7:** Plot of the intrinsic frequency $\omega$ as a function of $\rho$ and $\beta$. The dark region is where a 2 to 8 year recurrence period is observed. Parameter values are in Table 1.

We are interested in the parameter range when $\mathcal{R}_0 > 1$, that is, where the disease has recurring epidemics (see Figure 6(c)). When the $\beta$ and $\rho$ values are both very low, the intrinsic frequency is near zero, and so all populations fluctuate around their endemic equilibrium very slowly (see Figure 7). For $(\rho, \beta)$ values where the approximation is valid, the dominant outbreak period is 2 to 8 years, as seen by comparing the grey region in Figure 6(a) with Figure 7.

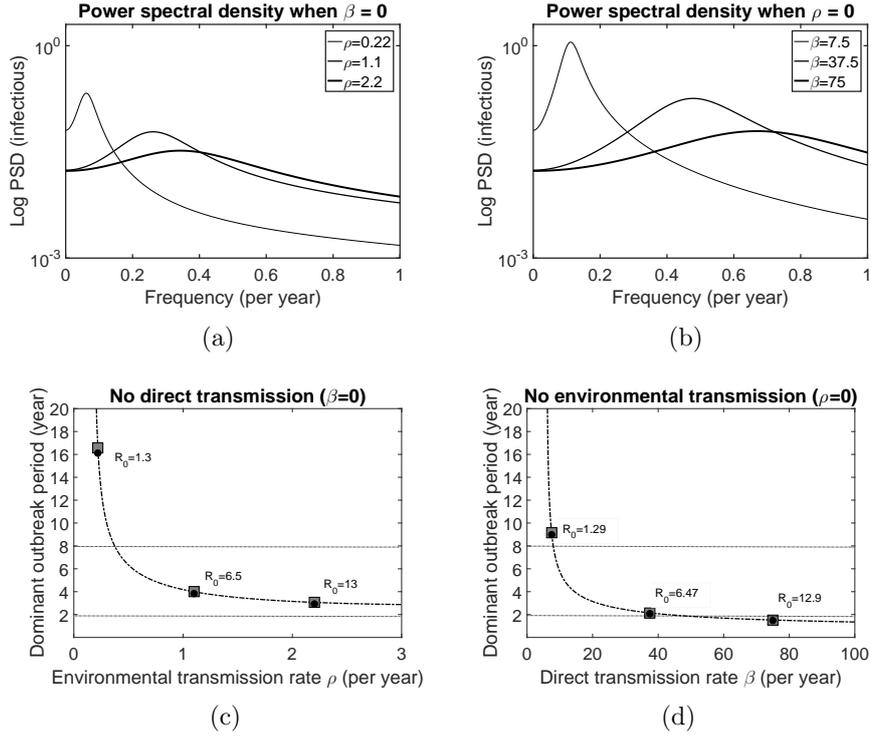

**Figure 8:** *Top panels:* Theoretical PSDs of the infectious population fluctuations $\xi_I(t)$ for (a) $\rho = 0.22, 1.1, 2.2$ when $\beta = 0$ and (b) $\beta = 7.5, 37.5, 75$ when $\rho = 0$. The linewidth of the PSD curves increases with the transmission parameter values. *Bottom panels:* Approximate dominant outbreak period, $2\pi/\omega$ with formula (69), as a function of (c) $\rho$ when $\beta = 0$ and (d) $\beta$ when $\rho = 0$. The square markers in (c) and (d) are located at $\rho = 0.22, 1.1, 2.2$ and $\beta = 7.5, 37.5, 7.5$, respectively. The black dots represent the exact dominant outbreak period obtained using the theoretical PSDs in the top panels. The black dots and square markers are indistinguishable from each other. The horizontal lines in (c) and (d) indicate the 2 to 8 year periods observed in actual prevalence data (See (Wang et al., 2012)). The corresponding $R_0$ values are also shown. Default values for all other parameters are given in Table 1.

We determine the outbreak periodicity for various values of either $\beta$ or $\rho$ alone by computing the theoretical PSD of the linear system (7) with (8) and (9) (Figure 8). When there is no direct transmission, the dominant frequency of the simulated outbreaks increases with the environmental transmission rate (Figure 8(a)). Similarly, the dominant frequency increases with the direct transmission rate in the absence of environmental transmission (Figure 8(b)).

In parallel with the increasing dominant frequency, the width of the PSD also increases with increasing environmental transmission rate. This means that, at low transmission rates, the outbreak pattern is more regular, and becomes more irregular as transmission rate increases. Consequently, when transmission rates are high, the timing of outbreaks is more difficult to predict.

Note here that we have chosen values of $\rho$ (when $\beta = 0$) and $\beta$ (when $\rho = 0$) that give roughly the same basic reproduction number $\mathcal{R}_0$, allowing for comparison between the theoretical PSDs. Consider the PSDs associated with $(\rho, \beta) = (0.22, 0)$ and $(\rho, \beta) = (0, 7.5)$ in Figure 8(b) and 8(a), respectively. Both parameter values correspond to $\mathcal{R}_0 \approx 1.3$, a value used by Wang et al. (2012) to define a boundary between disease persistence and stochastic extinction for their model. If we compare the PSDs between Figures 8(a) and 8(b), we find that, with the values in Table 1, the model predicts larger-valued PSDs in the case when environmental transmission is absent than when direct transmission is absent.

The corresponding dominant outbreak period is shown in Figure 8(c) and 8(d). Both dominant periods of the approximate and exact process are shown. The results are indistinguishable. We also observe in Figure 8(c) that, for $\beta = 0$, an increase in $\rho$ from 0.22 to 1.1 results in a decline of approximately 12 years in the outbreak period. However, for $\rho = 0$, an increase in $\beta$ from 7.5 to 37.5 only results in a decrease of approximately 7 years in the outbreak period (see Figure 8(d)). In other words, when the disease is epidemic but $\mathcal{R}_0$ is close to 1, increasing the environmental transmission can cause a greater drop in the outbreak period than increasing direct transmission. However, for larger transmission rates we find that the dominant outbreak period declines slowly as transmission rate increases and so the effects of changes in the individual transmission rates may be difficult to distinguish.

### 4.2.2 Typical intensity of outbreaks

For each $i = S, I, V$, the stationary standard deviation of $\xi_i^{app}(t)$ is

$$SSD_i = \sqrt{\frac{q_{i1}^2 \sigma_1^2}{2\zeta} + r_i^2 \frac{\tilde{\sigma}^2}{\lambda}} \qquad (27)$$

for $r_i^2 = q_{i2}^2 + q_{i3}^2$ where $q_{ij}$ are entries of the matrix $\mathbf{Q}$, e.g. (20), to determine the typical intensity of outbreaks. Here we show how the typical outbreak intensity is influenced by each type of transmission. In Figure 9, we display the plots of $SSD_i$'s of each process $\xi_i^{app}(t)$ as functions of $\beta$ and $\rho$.

In Figure 9, we have focused on the $(\rho, \beta)$ region that corresponds to $\mathcal{R}_0 > 1$ and $\lambda/\omega$ small, i.e. where the approximation method is valid. Typical amplitudes in simulated epidemics when $0 \leq \rho \leq 1$ and $0 \leq \beta \leq 100$ are depicted in Figure 9.

In Figure 9(b), we see that in general, fluctuation amplitudes are higher for larger values of $\beta$, the direct transmission rate. On the other hand, in Figure 9(a), the fluctuation amplitudes for the susceptibles are lower for larger $\beta$, and there is high sensitivity to $\beta$. Figure 9(c) shows an optimal region in $\beta$ for the fluctuation amplitude of virus especially for low $\rho$.

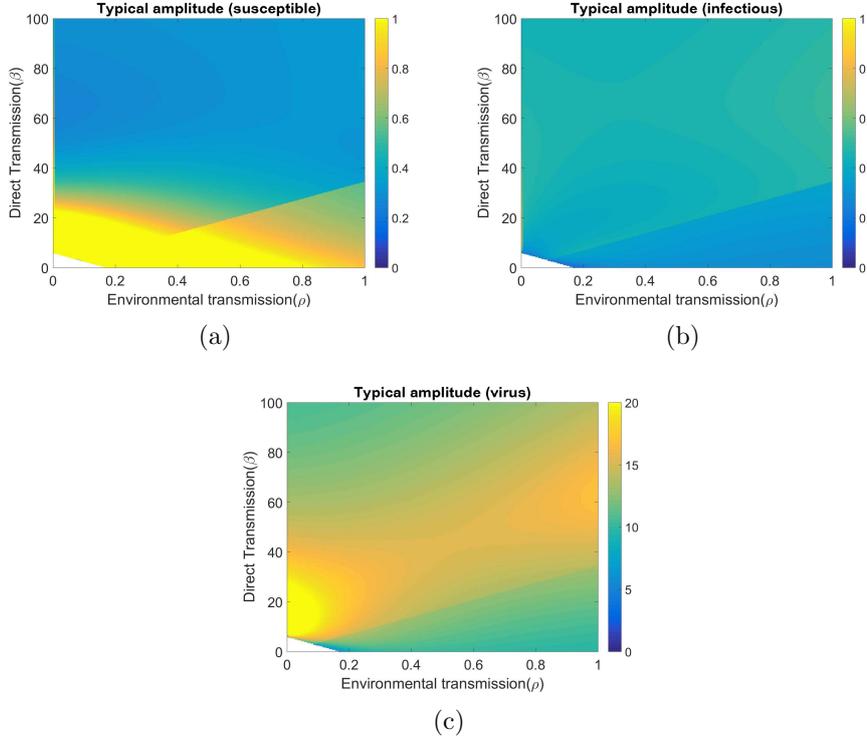

**Figure 9:** Typical amplitude (intensity) of the fluctuations in the proportion of (a) susceptible, (b) infectious, and (c) virus populations, measured by their stationary standard deviations (27) as a function of the environmental transmission rate $\rho$ and $\beta$. All other parameters are at the default values in Table 1.

## 5 Discussion

We have written the Wang et al. (2012) stochastic model for avian influenza including direct and environmental transmission routes as stochastic differential equations using the method of Kurtz (1978), following Greenwood and Gordillo (2009). Under large host and virus populations, the stochastic model approaches the deterministic system wherein, for $\mathcal{R}_0 > 1$, the endemic steady state is a stable focus. Our discussion of the avian flu model in Section 2 suggests that the disease can persist ($\mathcal{R}_0 > 1$) if either one of the two transmission routes is sufficiently strong. We have also shown via stochastic simulations that the model gives rise to noise sustained oscillation in the presence of either transmission route.

Our analysis allows us to conclude that the temporal pattern of epidemic recurrence is the sum of two processes: (1) a scaled OU process with long-term mean zero, and (2) the product of a rotation and a slowly varying standard OU process in two dimensions. This structure holds for any disease where the decay rate in the amplitude of successive epidemics is sufficiently slower than the frequency of recurrent epidemics. That is, as the deterministic process decays, there will be several noticeable epidemics before the system decays to near the steady state.

After linearisation, we study the stochastic path in three-dimensional space. The sample path behaves as an OU process that travels along the axis pointing in the direction of an eigenvector associated with the negative real eigenvalue, and cycles on the subspace spanned by the eigenvectors associated with the complex eigenvalues (See Appendix D).

We have shown that there is good agreement between the theoretical PSDs of the exact and approximate processes for each population type considered in the system. Although we observe small differences in the PSDs of the approximate and exact processes, we find that the two PSDs have closely matching dominant frequencies. The dominant frequency of the simulated epidemics is close to the intrinsic frequency $\omega$. We know from e.g. Greenwood et al. (2014) that the difference between the dominant and intrinsic frequencies is $\mathcal{O}(\lambda)$, where $\lambda$, which is small here, represents the decay rate of the deterministic solution. Based on the polar form of the approximate process (26), each population cycles at a frequency corresponding to the intrinsic frequency $\omega$ perturbed by a stochastic phase process $\varphi_{\lambda t}$.

We also notice that the PSDs of the exact process appear flat for low frequency and have a downward slope for high frequency, which are features of the PSD of a stationary OU process (Gardiner, 1986). This observation is in agreement with the approximate process we derived (23), which is a sum of an OU process (24), and the product of a rotation and a bi-variate standard OU process.

Previous studies (Breban and Drake, 2009; Rohani et al., 2009; Wang et al., 2012) of the recurrence of avian flu epidemics have focussed on identifying mechanisms that explain the multi-annual periodicity of the disease. Wang et al. (2012) claimed that environmental transmissibility is an important ingredient in explaining the 2 to 8 year period of avian flu. Using theoretical PSDs, they showed that for a fixed direct transmission rate, the outbreak period decreases as the environmental transmission rate is increased. Their results are based on the assumption that direct transmission is weak. However, according to Roche and Lebarbenchon (2009), the direct transmission rate $\beta$ can have a wide range of values. Our results show that the 2-8 year outbreak period can also be obtained chiefly as a result of direct transmission, and even in the absence of environmental transmission. Thus we conclude that both transmission rates are important factors in understanding the multi-year periodicity of disease outbreaks.

Our approach also allows us to obtain approximate values for the typical amplitude of the population fluctuations, because the stationary behaviour of the OU process is known. For the given avian flu parameter values, we find that the virus population fluctuations peak after the infectious population fluctuations peak. Typically, the phase lag is on the order of $0.3127/\omega = 0.31/0.8377 \approx 0.37$ of a year, i.e. approximately 4 months. This lag is reasonable as we have evidence that virus can be excreted by an infected bird for many days after infection (Alexander et al., 1986).

High amplitude epidemics arise when the direct transmission rate is high. In this case, the stochastic perturbation phase process changes slowly and so the dominant frequency of epidemics is comparatively regular.

Our analysis emphasizes that the interaction of stochasticity and transmission routes indeed plays an important role in determining outbreak periodicity and intensity. We suggest here that the recurrent pattern of avian flu outbreaks in itself is a result of noise amplification wherein its periodicity and amplitudes are influenced by either or both of the modes of transmission. The approach we introduced here could be used to perform a systematic study of other recurrent diseases.

**Competing interests.** We declare we have no competing interests.

**Funding.** Funding for this research was provided by NSERC (RGPIN-2016-05277) [RCT].

**Acknowledgement.** We gratefully acknowledge the BRAES institute at UBCO for supporting this work.

# 6 Appendices

# A Derivation of the avian flu SDE system

First, we define the probability of a jump or an increment $\Delta \vec{X}$ as

$$P(\Delta \vec{X} = \boldsymbol{\sigma}_{t+\Delta t} - \boldsymbol{\sigma}_t) = T(\boldsymbol{\sigma}'|\boldsymbol{\sigma})\Delta t.$$

Note that the increments of stochastic processes $S_t, I_t$ and $V_t$ are $\Delta S = S_{t+\Delta t} - S_t$, $\Delta I = I_{t+\Delta t} - I_t$, and $\Delta V = V_{t+\Delta t} - V_t$, respectively. Then, the expected values of the increments given the transition probabilities in Section 2.1 are

$$\begin{aligned} E[\Delta S] &= -\left(\beta \frac{S}{N} I + \rho S \frac{V}{N_V}\right) \Delta t + \mu(N - S - I)\Delta t + \mu I \Delta t, \\ E[\Delta I] &= \left(\beta \frac{S}{N} I + \rho S \frac{V}{N_V}\right) \Delta t - \mu I \Delta t - \gamma I \Delta t, \\ E[\Delta V] &= \tau I \Delta t + \delta V \Delta t - \eta V \Delta t. \end{aligned} \quad (28)$$

Now, each increment can be expressed as the expected value of the increment plus a sum of centred increments (Greenwood and Gordillo, 2009). Hence, we write the increments as:

$$\begin{aligned} \Delta S &= \left(-\beta \frac{S}{N} I - \rho S \frac{V}{N_V} + \mu(N - S - I) + \mu I\right) \Delta t - \Delta Z_1 + \Delta Z_2 + \Delta Z_3, \\ \Delta I &= \left(\beta \frac{S}{N} I + \rho S \frac{V}{N_V} - \mu I - \gamma I\right) \Delta t + \Delta Z_1 - \Delta Z_3 - \Delta Z_4, \\ \Delta V &= (\tau I + \delta V - \eta V) \Delta t + \Delta Z_5 - \Delta Z_6. \end{aligned} \quad (29)$$

Here the quantities $\Delta Z_i$ are conditionally centred Poisson increments with mean zero with conditional variances that are related to the transition rates. The Poisson increment $\Delta Z_1$ corresponding to infection of a susceptible individual has a conditional variance $\left(\beta \frac{S}{N} I + \rho S \frac{V}{N_V}\right) \Delta t$. The increments $\Delta Z_2$, and $\Delta Z_3$ corresponding to births in susceptible class (or deaths in recovered and infected class) respectively have conditional variances $\mu(N - S - I)\Delta t$, and $\mu I \Delta t$. On the other hand, the Poisson increment $\Delta Z_4$ that corresponds to recovery of an infected individual has a conditional variance equal to $\gamma I \Delta t$. Finally, the two increments corresponding to the replication and decay of viruses $\Delta Z_5$ and $\Delta Z_6$ must have conditional variances $(\tau I + \delta V)\Delta t$ and $\eta V \Delta t$, respectively. Divide (29) by $N$ and $N_V$ appropriately and take $\Delta t \to 0$ to obtain

$$
\begin{aligned}
dS &= \left(-\beta \frac{S}{N} I - \rho S \frac{V}{N_V} + \mu(N - S - I) + \mu I\right) dt - dZ_1 + dZ_2 + dZ_3, \\
dI &= \left(\beta \frac{S}{N} I + \rho S \frac{V}{N_V} - \mu I - \gamma I\right) dt + dZ_1 - dZ_3 - dZ_4, \\
dV &= (\tau I + \delta V - \eta V) dt + dZ_5 - dZ_6.
\end{aligned}
\tag{30}
$$

Suppose we replace the Poisson increments in (29) by multiples of Wiener increments, i.e. $\Delta Z_i \to g_i \Delta W_i$, with same standard deviations as the Poisson increments they replace. By doing the same limiting process $\Delta t \to 0$, we obtain the stochastic differential equations (SDE):

$$
\begin{aligned}
dS &= \left(-\beta \frac{S}{N} I - \rho S \frac{V}{N_V} + \mu(N - S - I) + \mu I\right) dt - g_1 dW_1 + g_2 dW_2 + g_3 dW_3, \\
dI &= \left(\beta \frac{S}{N} I + \rho S \frac{V}{N_V} - \mu I - \gamma I\right) dt + g_1 dW_1 - g_3 dW_3 - g_4 dW_4, \\
dV &= (\tau I + \delta V - \eta V) dt + g_5 dW_5 - g_6 dW_6,
\end{aligned}
\tag{31}
$$

where

$$
\begin{aligned}
g_1 &= \sqrt{\beta \frac{S}{N} I + \rho S \frac{V}{N_V}}, \qquad g_2 = \sqrt{\mu(N - S - I)}, \qquad g_3 = \sqrt{\mu I}, \\
g_4 &= \sqrt{\gamma I}, \qquad g_5 = \sqrt{\tau I + \delta V}, \qquad \text{and} \qquad g_6 = \sqrt{\eta V}.
\end{aligned}
\tag{32}
$$

Furthermore, we can re-write (31) by expressing the host and virus populations as proportions rather than absolute numbers, i.e.

$$
s = \frac{S}{N}, i = \frac{I}{N}, v = \frac{V}{N_V}, \text{and } k = \frac{N}{N_V}.
$$

The corresponding SDEs for the proportions of ducks and virus are then given by (2).

The approximation (31) is an example of a result of Kurtz (1978). An alternate approach is to use a van Kampen (Van Kampen, 1992) system-size expansion of the Kolmogorov (Master) equation, see e.g. in Baxendale and Greenwood (2011).

## B  Stochastic linearization

In matrix notation, (2) can be written as:

$$
d\mathbf{x} = \mathbf{F}(\mathbf{x}(t)) dt + \mathbf{D}\mathbf{G}(\mathbf{x}(t)) d\mathbf{W} \tag{33}
$$

where $\quad \mathbf{D} = diag(\frac{1}{\sqrt{N}}, \frac{1}{\sqrt{N}}, \frac{1}{\sqrt{N_V}}), \quad d\mathbf{W}(t) = (dW_1, dW_2, dW_3, dW_4, dW_5, dW_6)^T,$

and $\quad \mathbf{G}(\mathbf{x}(t)) = \begin{bmatrix} -G_1 & G_2 & G_3 & 0 & 0 & 0 \\ G_1 & 0 & -G_3 & -G_4 & 0 & 0 \\ 0 & 0 & 0 & 0 & G_5 & -G_6 \end{bmatrix}.$

Note that $\mathbf{x} = (s, i, v)$ which depends on $N$ and $N_V$ and $\lim_{N, N_V \to \infty} \mathbf{F}(\mathbf{x})$ is a vector whose components are the right-hand side of (4). It has been pointed out by Allen et al. (2008) that one can construct a stochastic system, which is the same in distribution such that all matrices in the diffusion term of (33) are square matrices whose sizes are equal to the dimension of $\mathbf{x}$, i.e. in this case, a matrix $\mathbf{C} \in \mathbb{R}^{3 \times 3}$ such that (33) would be equivalent in law to the stochastic system

$$d\tilde{\mathbf{x}} = \mathbf{F}(\tilde{\mathbf{x}}(t)) \, dt + \mathbf{D}\mathbf{C}(\tilde{\mathbf{x}}(t)) \, d\tilde{\mathbf{W}}. \tag{34}$$

The Wiener processes $\tilde{\mathbf{W}} \in \mathbb{R}^{3 \times 1}$ and $\mathbf{W} \in \mathbb{R}^{6 \times 1}$ both have independent terms. Moreover, the stochastic processes $\tilde{\mathbf{x}}$ in (34) are different from the originally defined stochastic processes found in (33) but it can be shown that their stochastic paths are the same. Thus, $\tilde{\mathbf{x}}$ can be replaced by the $\mathbb{R}^3$-valued stochastic process $\mathbf{x}$ that is considered originally. Matrices $\mathbf{G}$ and $\mathbf{C}$ are related through the $3 \times 3$ matrix $\mathbf{V}$, where $\mathbf{V} = \mathbf{G}\mathbf{G}^\intercal$ and $\mathbf{C} = \mathbf{V}^{1/2}$. An explicit computation of $\mathbf{V}$ confirms that it is the general form for the noise covariance matrix $\mathbf{B}$ that was described in Wang et al. (2012). In other words,

$$\mathbf{V}(\mathbf{x}, t) = \begin{bmatrix} \beta si + \rho sv + \mu(1-s) & -\beta si - \rho sv - \mu i & 0 \\ -\beta si - \rho sv - \mu i & \beta si + \rho sv + (\mu + \gamma)i & 0 \\ 0 & 0 & k\tau i + \delta v + \eta v \end{bmatrix}. \tag{35}$$

Letting $N, N_V \to \infty$ so that $s \to \phi_1, i \to \phi_2,$ and $v \to \psi$ and $t \to \infty$ we have $\phi_1 \to \phi_1^*, \phi_2 \to \phi_2^*,$ and $\psi \to \psi^*$ implies that $\lim_{N, N_V, t \to \infty} \mathbf{V} = \mathbf{B}$ which is a constant matrix whose entries are displayed as follows where $\mathbf{x_{eq}} \equiv (\phi_1 = \phi_1^*, \phi_2 = \phi_2^*, \psi = \psi^*)$, the equilibrium state of the deterministic system:

$$\mathbf{B} = \begin{bmatrix} B_{11} & B_{12} & 0 \\ B_{21} & B_{22} & 0 \\ 0 & 0 & B_{33} \end{bmatrix}, \qquad \text{where}$$
$$B_{11} = \beta\phi_1^*\phi_2^* + \rho\phi_1^*\psi^* + \mu(1-\phi_1^*),$$
$$B_{12} = B_{21} = -\beta\phi_1^*\phi_2^* - \rho\phi_1^*\psi^* - \mu\phi_2^*, \tag{36}$$
$$B_{22} = \beta\phi_1^*\phi_2^* + \rho\phi_1^*\psi^* + (\mu + \gamma)\phi_2^*, \qquad \text{and}$$
$$B_{33} = \kappa\tau\phi_2^* + \delta\psi^* + \eta\psi^*.$$

It remains to show that the set of Langevin equations obtained by Wang et al. (2012) can be constructed from the linear stochastic differential equations (the tilde in (34) is dropped for brevity)

$$d\mathbf{x} = \mathbf{F}(\mathbf{x}(t)) \, dt + \mathbf{D}\mathbf{C}(\mathbf{x}(t)) \, d\mathbf{W}. \tag{37}$$

Recall that the diagonal matrix $\mathbf{D}$ is given in (33) and $\mathbf{C} = \mathbf{V}^{1/2}$ where the entries of $\mathbf{V}$ is described in (35). The system (37) with the stochastic term generates the deterministic process (4), since (37) becomes (4) as $N, N_V \to \infty$ which means that this term describes the average dynamics of the processes. On the other hand, the second term is referred to as the diffusion term. It represents the variation from the average dynamics, the $\mathcal{O}(N^{-1/2})$ fluctuations of $\mathbf{x}(t)$ away from the deterministic process. The diffusion term prevents a damped system from settling to an equilibrium state.

We linearize (37) using the substitution $\mathbf{x}(t) = \mathbf{x_{eq}} + \mathbf{D}\boldsymbol{\xi}(t)$ and obtain

$$\mathbf{D}d\boldsymbol{\xi} = \mathbf{F}(\mathbf{x_{eq}})\ dt + \mathbf{D}\mathbf{J}(\mathbf{x_{eq}})\boldsymbol{\xi}\ dt + \mathbf{D}\mathbf{C}(\mathbf{x_{eq}})\ d\mathbf{W}. \tag{38}$$

The Jacobian of $\mathbf{F}(\mathbf{x})$ evaluated at $\mathbf{x_{eq}}$ is denoted by $\mathbf{J}(\mathbf{x_{eq}})$. Now, $\mathbf{F}(\mathbf{x_{eq}}) = \mathbf{0}$ and so simplifies (38), after pre-multiplying by $\mathbf{D}^{-1}$, to

$$d\boldsymbol{\xi} = \mathbf{J}(\mathbf{x_{eq}})\boldsymbol{\xi}\ dt + \mathbf{C}(\mathbf{x_{eq}})\ d\mathbf{W}. \tag{39}$$

Eq. (39) is the Langevin (i.e. stochastic) equation in Wang et al. (2012)(See Eq.6) written in slightly different form. In particular, the two equations would be equivalent if we divide (39) by $dt$ and denote $\mathbf{A} = \mathbf{J}(\mathbf{x_{eq}})$ and represent the diffusion term as $\boldsymbol{\zeta}(t)$, i.e. Gaussian white noise with correlation function $\langle \boldsymbol{\zeta}(t), \boldsymbol{\zeta}(t')^T \rangle = \mathbf{B}\delta(t-t')$. In (7), we have $\mathbf{A_0} = \mathbf{J}(\mathbf{x_{eq}})$ and $\mathbf{C_0} = \mathbf{C}(\mathbf{x_{eq}})$.

## C    Approximate solution for linear diffusion equations in three dimensions

We follow Baxendale and Greenwood (2011) to derive the approximate solution for our example where the diffusion processes have values in $\mathbb{R}^3$.

Consider the stochastic system

$$d\boldsymbol{\xi} = \mathbf{A_0}\boldsymbol{\xi}\ dt + \mathbf{C_0}\ d\mathbf{W}, \qquad \boldsymbol{\xi}(t), \mathbf{W}(t) \in \mathbb{R}^3, \mathbf{A_0}, \mathbf{C_0} \in \mathbb{R}^{3\times 3}. \tag{40}$$

where,

$$\mathbf{A_0} = \begin{bmatrix} -\beta\phi_2^* - \rho\psi^* - \mu & -\beta\phi_1^* & -\rho\phi_1^* \\ -\beta\phi_2^* & \beta\phi_1^* - \mu - \gamma & \rho\phi_1^* \\ 0 & \kappa\tau & \delta - \eta \end{bmatrix}, \tag{41}$$

and

$$\mathbf{C_0} = \begin{bmatrix} \beta\phi_1^*\phi_2^* + \rho\phi_1^*\psi^* + \mu(1-\phi_1^*) & -\beta\phi_1^*\phi_2^* - \rho\phi_1^*\psi^* - \mu\phi_2^* & 0 \\ -\beta\phi_1^*\phi_2^* - \rho\phi_1^*\psi^* - \mu\phi_2^* & \beta\phi_1^*\phi_2^* + \rho\phi_1^*\psi^* + (\mu+\gamma)\phi_2^* & 0 \\ 0 & 0 & \kappa\tau\phi_2^* + \delta\psi^* + \eta\psi^* \end{bmatrix}^{1/2}. \tag{42}$$

Here $\mathbf{W}(t)$ contains independent Wiener processes (or Brownian motion).

Suppose that $\mathbf{A_0}$ has eigenvalues $-\zeta$ and $-\lambda \pm i\omega$ for $\zeta, \lambda, \omega \in \mathbb{R}^+$. One can find a matrix $\mathbf{Q} \in \mathbb{R}^{3\times 3}$ such that

$$\mathbf{Q}^{-1}\mathbf{A_0}\mathbf{Q} = \boldsymbol{\Lambda} \equiv \begin{bmatrix} -\zeta & 0 & 0 \\ 0 & -\lambda & \omega \\ 0 & -\omega & -\lambda \end{bmatrix}. \tag{43}$$

The matrix $\mathbf{\Lambda}$ is called the real block diagonal form of the eigenvalue matrix of $\mathbf{A_0}$ so it follows that $\mathbf{Q}$ is the real block diagonal form of the associated matrix of eigenvectors. By pre-multiplying (40) with $\mathbf{Q}^{-1}$ and using the substitution $\mathbf{y}(t) = \mathbf{Q}^{-1}\boldsymbol{\xi}(t)$, we have

$$d\mathbf{y} = \mathbf{\Lambda}\mathbf{y}\ dt + \mathbf{Q}^{-1}\mathbf{C_0}\ d\mathbf{W}. \tag{44}$$

Let $\mathbf{\Sigma} = \mathbf{Q}^{-1}\mathbf{C_0}$ and denote $\mathbf{\Sigma}_{\bullet j}$ and $\mathbf{\Sigma}_{i\bullet}$ as its $j$th column vector and $i$th row vector, respectively. With $\mathbf{y} = [y_1, y_2, y_3]^\intercal$, we write (44) as

$$dy_1 = -\zeta y_1\ dt + \mathbf{\Sigma}_{1\bullet}\ d\mathbf{W}, \tag{45a}$$

$$d\tilde{\mathbf{y}} = \tilde{\mathbf{\Lambda}}\tilde{\mathbf{y}}\ dt + \tilde{\mathbf{\Sigma}}\ d\mathbf{W}, \tag{45b}$$

where $\quad \tilde{\mathbf{y}} = [y_2, y_3]^\intercal$, $\tilde{\mathbf{\Lambda}} = \begin{bmatrix} -\lambda & \omega \\ -\omega & -\lambda \end{bmatrix}$, and $\tilde{\mathbf{\Sigma}} = [\mathbf{\Sigma}_{2\bullet}, \mathbf{\Sigma}_{3\bullet}]^\intercal$.

Now, using the result of Allen et al. (2008), we find that the SDE (45a) is equivalent to

$$dy_1 = -\zeta y_1\ dt + \sigma_1\ dW_1, \tag{46}$$

where $\sigma_1^2 = \mathbf{\Sigma_{1\bullet}}\mathbf{\Sigma_{1\bullet}^\intercal}$ is the variance of the stationary distribution of $y_1(t)$ and $W_1(t)$ is a one-dimensional Wiener process. It is apparent that (46) describes an Ornstein-Uhlenbeck process (Uhlenbeck and Ornstein, 1930) in one dimension with a stationary variance $\sigma_1^2/2\zeta$. The square-root of this stationary variance corresponds to the standard deviation typically observed in the process $y_1(t)$.

On the other hand, (45b) is equivalent to:

$$d\tilde{\mathbf{y}} = \tilde{\mathbf{\Lambda}}\tilde{\mathbf{y}}\ dt + \tilde{\mathbf{C}}\ d\tilde{\mathbf{W}}, \tag{47}$$

where $\tilde{\mathbf{C}} = (\tilde{\mathbf{\Sigma}}\tilde{\mathbf{\Sigma}}^\intercal)^{1/2}$ and $\tilde{\mathbf{W}}(t)$ is a two-dimensional Wiener process. The approximate solution of (47) is related to a two-dimensional Ornstein-Uhlenbeck process as proven by Baxendale and Greenwood (2011). The approximation is reasonable under the assumption that $\lambda \ll \omega$. Thus, if it is assumed that $\lambda \ll \omega$ then by the theorem of Baxendale and Greenwood (2011), the approximate solution for $\tilde{\mathbf{y}}$ is:

$$\tilde{\mathbf{y}} \approx \tilde{\mathbf{y}}^{\mathbf{app}} = \frac{\tilde{\sigma}}{\sqrt{\lambda}}\mathbf{R}_{-\omega t}\mathbf{S}_{\lambda t}, \tag{48}$$

where

$$\tilde{\sigma}^2 = \frac{1}{2}Tr(\tilde{\mathbf{C}}\tilde{\mathbf{C}}^\intercal). \tag{49}$$

Thus,

$$\boldsymbol{\xi} \approx \boldsymbol{\xi}^{app} \equiv \mathbf{Q}\mathbf{y}^{app} = y_1\mathbf{Q}_{\bullet 1} + y_2^{app}\mathbf{Q}_{\bullet 2} + y_3^{app}\mathbf{Q}_{\bullet 3} \\ = y_1\mathbf{Q}_{\bullet 1} + [\mathbf{Q}_{\bullet 2}, \mathbf{Q}_{\bullet 3}]\tilde{\mathbf{y}}^{\mathbf{app}}. \tag{50}$$

More precisely,

$$\boldsymbol{\xi}^{app}(t) = y_1(t)\mathbf{Q}_{\bullet 1} + \frac{\tilde{\sigma}}{\sqrt{\lambda}}[\mathbf{Q}_{\bullet 2}, \mathbf{Q}_{\bullet 3}]\mathbf{R}_{-\omega t}\mathbf{S}_{\lambda t}. \tag{51}$$

Now, we know that in polar coordinates

$$\mathbf{R}_{-\omega t}\mathbf{S}_{\lambda t} = \begin{bmatrix} \cos\omega t & \sin\omega t \\ -\sin\omega t & \cos\omega t \end{bmatrix} \begin{bmatrix} S_1(\lambda t) \\ S_2(\lambda t) \end{bmatrix} = \begin{bmatrix} S_1(\lambda t)\cos\omega t + S_2(\lambda t)\sin\omega t \\ -S_1(\lambda t)\sin\omega t + S_2(\lambda t)\cos\omega t \end{bmatrix}. \qquad (52)$$

Using the formula $x\cos\theta + y\sin\theta = z\cos(\theta - \varphi)$ where $z = |x+iy|$ and $\varphi = \arg(x+iy)$ for $x, y \in \mathbb{R}$ and $i = \sqrt{-1}$, we then write $S_1(\lambda t) = z(\lambda t)\cos\varphi(\lambda t)$ and $S_2(\lambda t) = z(\lambda t)\sin\varphi(\lambda t)$ with $z(\lambda t) = \sqrt{S_1^2 + S_2^2} \equiv |\mathbf{S}(\lambda t)|$ and $\varphi(\lambda t) = \tan^{-1}(S_2/S_1)$ to obtain

$$\mathbf{R}_{-\omega t}\mathbf{S}_{\lambda t} = z(\lambda t) \begin{bmatrix} \cos(\varphi(\lambda t) - \omega t) \\ \sin(\varphi(\lambda t) - \omega t) \end{bmatrix} \equiv |\mathbf{S}(\lambda t)| \begin{bmatrix} \cos(\varphi_{\lambda t} - \omega t) \\ \sin(\varphi_{\lambda t} - \omega t) \end{bmatrix}. \qquad (53)$$

Applying (53) to the second term of (51) yields

$$\begin{bmatrix} \xi_1^{app}(t) \\ \xi_2^{app}(t) \\ \xi_3^{app}(t) \end{bmatrix} = y_1(t) \begin{bmatrix} q_{11} \\ q_{21} \\ q_{31} \end{bmatrix} + \frac{\tilde{\sigma}}{\sqrt{\lambda}}|\mathbf{S}(\lambda t)| \begin{bmatrix} q_{12}\cos(\varphi_{\lambda t} - \omega t) + q_{13}\sin(\varphi_{\lambda t} - \omega t) \\ q_{22}\cos(\varphi_{\lambda t} - \omega t) + q_{23}\sin(\varphi_{\lambda t} - \omega t) \\ q_{32}\cos(\varphi_{\lambda t} - \omega t) + q_{33}\sin(\varphi_{\lambda t} - \omega t) \end{bmatrix}, \qquad (54)$$

where $\mathbf{Q} = [q_{ij}]$.

We define $q_{i2} = r_i\cos\theta_i$ and $q_{i3} = r_i\sin\theta_i$ where $r_i = \sqrt{q_{i2}^2 + q_{i3}^2}$ and $\theta_i = \tan^{-1}(q_{i3}/q_{i2})$ so that the approximate fluctuation of each component takes the form:

$$\xi_i^{app}(t) = q_{i1}y_1(t) + \frac{\tilde{\sigma}}{\sqrt{\lambda}}|\mathbf{S}(\lambda t)|r_i\cos(\varphi_{\lambda t} - \omega t - \theta_i). \qquad (55)$$

The polar form of the approximation reveals that each model component fluctuates according to a combination of a univariate and bi-variate OU processes. The first term of the approximation contains a one-dimensional OU process weighted by a scalar determined from the transformation matrix $\mathbf{Q}$ while the second term contains the two-dimensional OU process that varies slowly and $\lambda t$ is a quantity that influences the radius and phase of the circular path. The stationary variance of $\xi_i^{app}(t)$ is the sum of the stationary variance of each term in (55). This means that the long-term variance of a fluctuation is

$$\frac{q_{i1}^2\sigma_1^2}{2\zeta} + r_i^2\frac{\tilde{\sigma}^2}{\lambda} \qquad (56)$$

Hence, the typical magnitude of $\xi_i^{app}(t)$, i.e. stationary standard deviation, is

$$SSD_i = \sqrt{\frac{q_{i1}^2\sigma_1^2}{2\zeta} + r_i^2\frac{\tilde{\sigma}^2}{\lambda}}. \qquad (57)$$

Note that the fluctuation of each component $i$ has a constant phase shift $\theta_i$, which is useful in computing phase differences between disease components.

# D  Additional insight from the approximation on the interaction of disease components

In Section 4.1, we showed that for the given set of avian flu parameter values, the system exhibits noise-sustained oscillations which can be viewed as a sum of two processes given by (23): (i) a process proportional to the one-dimensional OU process and (ii) a process proportional to the product of a rotation matrix and a standard OU process.

We describe here, using (11), the behaviour of the sample path in three-dimensional space. The first term of (11) means that sample path behaves as an Ornstein-Uhlenbeck process $y_1(t)$ that travels along the axis that points to the direction of $\mathbf{Q}_{\bullet 1}$, i.e. the eigenvector associated to $-\zeta$. In addition, the second term of (11) implies that the sample path cycles on the subspace spanned by the last two column vectors of the transformation matrix $\mathbf{Q}$, i.e. eigenvectors of the eigenvalues $-\lambda \pm i\omega$. This subspace contains a plane whose equation (see (62) in Appendix E for general formulation), for our chosen set of parameters, is given by:

$$\xi_S - 19.376\xi_I + 0.5814\xi_V = 0. \tag{58}$$

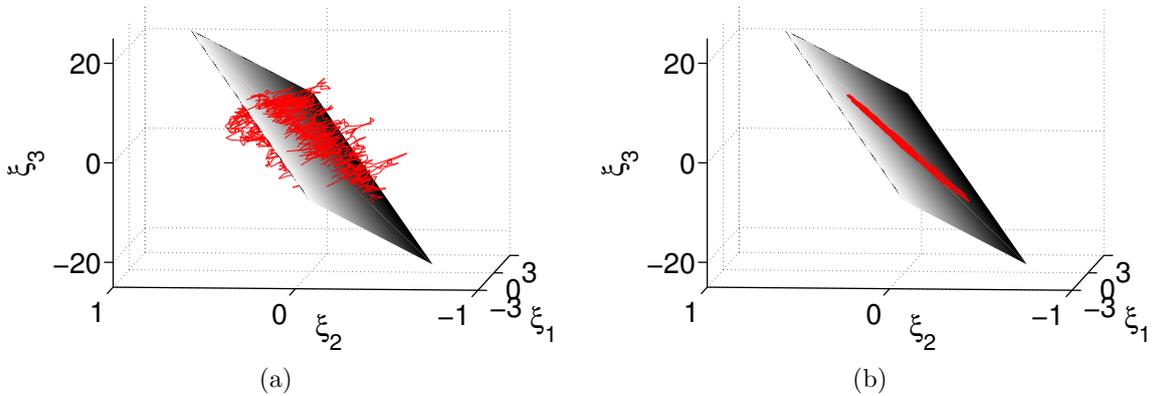

**Figure 10:** (Colour online) A sample path of the approximate fluctuations given by (23) when the first term is (a) not set to zero and (b) set to zero. The grey region is the plane (58) that lies in the subspace spanned by the eigenvectors of $-0.3091 \pm 0.8377i$.

Figure 10 shows the plane (58) and a realization of the stochastic simulation of (23). In Figure 10(a), we observe that the sample path lies chiefly on or near the plane (58). However, if we neglect the first term of (23), the dynamics of the fluctuations lie entirely on this plane (see Figure 10(b)). Thus, the portion of the sample path that departs from the plane is clearly due to the one-dimensional OU process whereas the second term constrains the sample path to move within the plane.

From (23), we know that the stationary standard deviation of $y_1(t)$ is 1.63 which is small compared to the constant $\tilde{\sigma}/\sqrt{\lambda} = 10.21$ that appears in the second term of the equation. Therefore, we can neglect the first term of (23) and show that avian flu epidemics cycle on the plane (58). This can be achieved mathematically when the magnitude of the real eigenvalue $\zeta$ is larger than $\lambda$, which means that the approximate process approaches the hyperplane in fast manner. In Figure 11, we compute the magnitude $\zeta$ over combinations of $\beta$ and $\rho$ values and found that the epidemic cycles occur primarily in the hyperplane when $\beta$ is below 100 and $\rho$ is high, i.e. where $\zeta$ is larger than $\lambda$.

The fact that avian flu dynamics could primarily occur in the plane suggest that under certain conditions for each transmission route, one can project the avian flu system (7) onto the plane (58) and so simplify the analysis. For instance, using the equation of the plane, we can write one component in terms of the other and convert the three-dimensional linear avian flu SDE system (7) into a two-dimensional one. The possibility of modelling recurrent avian flu epidemics as stochastic system in two dimensions must therefore be explored.

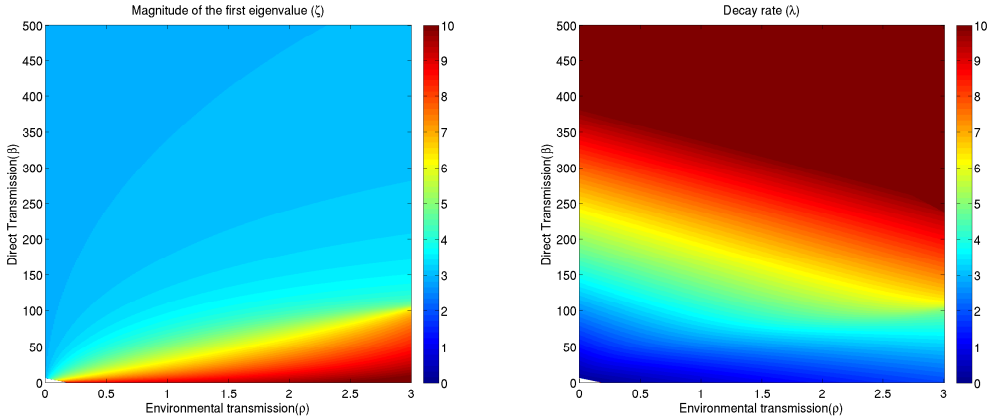

**Figure 11:** (Colour online) Plot of $\zeta$ (left panel) and $\lambda$ (right panel) as functions of $\beta$ and $\rho$. The white region is where $R_0 < 1$, i.e. noise-sustained oscillations cannot be observed here. Default parameter values are in Table 1.

# E The subspace where the cycling takes place

For the case when the stationary standard deviation (s.s.d.) of the second term is very large compared to the s.s.d. of the first term in our approximation, the first term of (51) is negligible and we expect the stochastic path to primarily lie in a plane, i.e. a subspace of $\mathbb{R}^3$, spanned by the last two column vectors of $\mathbf{Q}$ ($\mathbf{Q_{\bullet 2}}$ and $\mathbf{Q_{\bullet 3}}$). Here we show a general way for computing the equation of this plane.

The sample path defined by the fluctuations $\xi_i(t)$ is centred at $(0,0,0)$ and so the equation of the plane should take the form:

$$a_1\xi_1 + a_2\xi_2 + a_3\xi_3 = 0 \qquad (59)$$

We know that the vectors $\mathbf{Q_{\bullet 2}}$ and $\mathbf{Q_{\bullet 3}}$ span the plane, hence must satisfy (59). Therefore,

$$[a_1, a_2, a_3] \cdot [\mathbf{Q_{\bullet 2}}, \mathbf{Q_{\bullet 3}}] = \mathbf{0}. \qquad (60)$$

By Gaussian elimination or by manipulating the explicit form of the linear system, we can eliminate $a_3$ in (60) and a little algebra turns (60) into a simpler equation,

$$\det(\mathbf{M_1})a_1 + \det(\mathbf{M_2})a_2 = 0 \qquad \text{where} \qquad \mathbf{M_1} = \begin{bmatrix} q_{12} & q_{13} \\ q_{32} & q_{33} \end{bmatrix} \text{ and } \mathbf{M_2} = \begin{bmatrix} q_{22} & q_{23} \\ q_{32} & q_{33} \end{bmatrix}. \qquad (61)$$

Now we require $\det(\mathbf{M_2}) \neq 0$ so that $a_2 = -\dfrac{\det(\mathbf{M_1})}{\det(\mathbf{M_2})} a_1$, which gives $a_3 = \dfrac{a_1}{q_{32}} \left( \dfrac{\det(\mathbf{M_1})}{\det(\mathbf{M_2})} q_{22} - q_{12} \right)$.
Therefore, assuming that $a_1 \neq 0$, the desired equation of the plane is

$$\xi_1 - \frac{\det(\mathbf{M_1})}{\det(\mathbf{M_2})} \xi_2 + \frac{1}{q_{32}} \left( \frac{\det(\mathbf{M_1})}{\det(\mathbf{M_2})} q_{22} - q_{12} \right) \xi_3 = 0. \tag{62}$$

# F  Derivation of the explicit form of the mean-field eigenvalues

Our starting point is the Jacobian evaluated at the stable endemic equilibrium point (Wang et al., 2012), i.e.,

$$\mathbf{J}(\mathbf{x_{eq}}) = \begin{bmatrix} -\beta\phi_2^* - \rho\psi^* - \mu & -\beta\phi_1^* & -\rho\phi_1^* \\ -\beta\phi_2^* & \beta\phi_1^* - \mu - \gamma & \rho\phi_1^* \\ 0 & \kappa\tau & \delta - \eta \end{bmatrix}, \tag{63}$$

where $\phi_1^* = \dfrac{1}{\mathcal{R}_0}$, $\phi_2^* = \dfrac{\mu}{\mu+\gamma}\left(1 - \dfrac{1}{\mathcal{R}_0}\right)$, and $\psi^* = \dfrac{\kappa\mu\tau}{(\eta-\delta)(\mu+\gamma)}\left(1 - \dfrac{1}{\mathcal{R}_0}\right)$ for the basic reproduction number

$$\mathcal{R}_0 = \frac{\beta}{\mu+\gamma} + \frac{\kappa\rho\tau}{(\eta-\delta)(\mu+\gamma)}.$$

The condition $\mathcal{R}_0 > 1$ must be satisfied for the non-trivial steady-state $\mathbf{x_{eq}} = (\phi_1^*, \phi_2^*, \psi^*)$ to exist.

The eigenvalues of $\mathbf{J}(\mathbf{x_{eq}})$ determine the local dynamics of the deterministic SIR-V system close to the non-trivial equilibrium point $\mathbf{x_{eq}}$. Now, denote the eigenvalues of $\mathbf{J}(\mathbf{x_{eq}})$ as $\nu$. It follows that $|\mathbf{J}(\mathbf{x_e q}) - \nu I| = 0$ gives rise to a cubic polynomial of the form

$$\nu^3 - a\nu^2 - b\nu - c = 0, \tag{64}$$

where:

$$\begin{aligned} a &= (\delta - \eta) - \mu\mathcal{R}_0 - \gamma - \mu + \beta/\mathcal{R}_0 \\ b &= -\mu(\eta - \delta + \gamma + \mu)\mathcal{R}_0 + \mu\beta/\mathcal{R}_0 \\ c &= -\mu(\eta - \delta)(\gamma + \mu)(\mathcal{R}_0 - 1). \end{aligned} \tag{65}$$

Equations (64) and (65) in fact appeared in the Appendix section of Wang et al. (2012), where it was proven that the endemic equilibrium is stable. Now, substitute $\nu = y + \dfrac{a}{3}$ to yield the normal form transformation,

$$y^3 + py + q = 0 \quad \text{where} \quad p = \frac{1}{3}(-a^2 - 3b) \text{ and } q = \frac{1}{27}(-2a^3 - 9ab - 27c). \tag{66}$$

This method is also known as Vieta's subsitution Connor (1956). Equation (66) has been well-studied and has known solutions in general form:

$$\begin{aligned} y_1 &= Y_+ + Y_-, \\ y_2 &= -\frac{1}{2}(Y_+ + Y_-) + i\frac{\sqrt{3}}{2}(Y_+ - Y_-), \\ y_3 &= -\frac{1}{2}(Y_+ + Y_-) - i\frac{\sqrt{3}}{2}(Y_+ - Y_-), \end{aligned} \tag{67}$$

where: $Y_\pm = \left(-\dfrac{q}{2} \pm \sqrt{\dfrac{q^2}{4} + \dfrac{p^3}{27}}\right)^{1/3}$ and $i = \sqrt{-1}$.

We are interested in the case when all three roots exist with two of them being complex conjugates. This is satisfied by assuming that $\dfrac{q^2}{4} + \dfrac{p^3}{27} > 0$.

By back substitution, the solutions of (64) are:

$$\begin{aligned}
\nu_1 &= Y_+ + Y_- + \frac{a}{3}, \\
\nu_2 &= -\frac{1}{2}(Y_+ + Y_-) + \frac{a}{3} + i\frac{\sqrt{3}}{2}(Y_+ - Y_-), \\
\nu_3 &= -\frac{1}{2}(Y_+ + Y_-) + \frac{a}{3} - i\frac{\sqrt{3}}{2}(Y_+ - Y_-).
\end{aligned} \qquad (68)$$

The eigenvalues $\nu_2$ and $\nu_3$ are conjugate pairs whose real part is negative as confirmed by Wang et al. (2012). The magnitude of the real and imaginary part corresponds to the decay rate $\lambda$ and the intrinsic frequency $\omega$, respectively, of the deterministic system linearized at the endemic equilibrium state. Therefore,

$$\begin{aligned}
\lambda &= \left| -\frac{1}{2}(Y_+ + Y_-) + \frac{a}{3} \right|, \\
\omega &= \left| \frac{\sqrt{3}}{2}(Y_+ - Y_-) \right|.
\end{aligned} \qquad (69)$$

Using avian flu parameters in Table 1, we can write $\lambda$ and $\omega$ in terms of $\beta$ and $\rho$, as follows:

$$\lambda(\rho, \beta) \approx \Big| -2.9 - 0.0172\beta - 0.5945\rho + \frac{\beta}{0.5172\beta + 17.84\rho} \\ - 0.5\left(\sqrt[3]{\mathsf{F}(\rho,\beta) + \mathsf{G}(\rho\beta)} + \sqrt[3]{\mathsf{F}(\rho,\beta) - \mathsf{G}(\rho,\beta)}\right)\Big|, \qquad (70)$$

$$\omega(\rho,\beta) \approx \frac{\sqrt{3}}{2}\Big| -\sqrt[3]{\mathsf{F}(\rho,\beta) + \mathsf{G}(\rho,\beta)} + \sqrt[3]{\mathsf{F}(\rho,\beta) - \mathsf{G}(\rho,\beta)}\Big|, \qquad (71)$$

where

$$\begin{aligned}
\mathsf{F}(\rho,\beta) &= \frac{-0.5 P_1(\rho,\beta)}{(0.1724\beta + 5.945\rho)^3}, \\
\mathsf{G}(\rho,\beta) &= 0.56\sqrt{\frac{81 P_1(\rho,\beta)^2}{(0.1724\beta + 5.945\rho)^6} + \frac{12 P_2(\rho,\beta)^3}{(0.1724\beta + 5.945\rho)^6}}, \\
P_1(\rho,\beta) &= \sum_{i=0}^{6}\sum_{j=0}^{6} \mathsf{M}_{i+1,j+1} \beta^i \rho^j, \\
P_2(\rho,\beta) &= \sum_{i=0}^{6}\sum_{j=0}^{6} \mathsf{N}_{i+1,j+1} \beta^i \rho^j.
\end{aligned} \qquad (72)$$

Here
$$\mathsf{M} = \begin{bmatrix} 0 & 0 & 0 & 9190 & 3152 & -646 & 88.3 \\ 0 & 0 & 236 & 311 & -119 & 15.4 & 0 \\ 0 & 1.39 & 13.5 & -8.33 & 1.11 & 0 & 0 \\ -0.008 & 0.306 & -0.284 & 0.043 & 0 & 0 & 0 \\ 0.003 & -0.005 & 0.001 & 0 & 0 & 0 & 0 \\ 0 & 0 & 0 & 0 & 0 & 0 & 0 \\ 0 & 0 & 0 & 0 & 0 & 0 & 0 \end{bmatrix} \tag{73}$$

and
$$\mathsf{N} = \begin{bmatrix} 0 & 0 & -892 & 183 & -37.5 & 0 & 0 \\ 0 & -19 & 23 & -4.35 & 0 & 0 & 0 \\ -0.135 & 0.871 & -0.189 & 0 & 0 & 0 & 0 \\ 0.01 & -0.004 & 0 & 0 & 0 & 0 & 0 \\ 0 & 0 & 0 & 0 & 0 & 0 & 0 \\ 0 & 0 & 0 & 0 & 0 & 0 & 0 \\ 0 & 0 & 0 & 0 & 0 & 0 & 0 \end{bmatrix}. \tag{74}$$

Additionally, the first eigenvalue $\nu_1 < 0$ (eigenvalue with largest negative real part) and so the decay rate of the OU process $y_1(t)$ in (11) is $\zeta = |\nu_1|$.